\documentclass{elsart}

\usepackage{natbib}
\usepackage{graphicx}
\usepackage{amssymb}

\def\eq#1{\begin{equation} #1 \end{equation}}
\def\E#1{\hbox{$10^{#1}$}}
\def\about {\hbox{$\sim$}}
\def\la    {\hbox{$\lesssim$}}
\def\ga    {\hbox{$\gtrsim$}}
\def\x     {\hbox{$\times$}}
\def\mic   {\hbox{$\mu$m}}
\def\deg   {\hbox{$^\circ$}}

\def\LEdd   {\hbox{$L_{\rm Edd}$}}
\def\Mo     {\hbox{$M_{\odot}$}}
\def\tV    {\hbox{$\tau_{\rm V}$}}

\def\IC    {\hbox{$I^{\rm C}_{\lambda}$}}
\def\nc    {\hbox{$n_{\rm C}$}}
\def\Ac    {\hbox{$A_{\rm c}$}}
\def\Vc    {\hbox{$V_{\rm c}$}}
\def\N     {\hbox{$\cal N$}}
\def\No    {\hbox{${\cal N}_0$}}

\def\FAGN  {\hbox{$F_{\rm AGN}$}}
\def\Fcont {\hbox{$F_{\rm c,\lambda}$}}
\def\Pesc  {\hbox{$P_{\rm esc}$}}
\def\Ntor   {\hbox{$N_{\rm torus}^{\rm (eq)}$}}
\def\Mtor   {\hbox{$M_{\rm torus}$}}
\def\ntot   {\hbox{$n_{\rm tot}$}}
\def\Rd     {\hbox{$R_{\rm d}$}}
\def\Ro     {\hbox{$R_{\rm o}$}}
\def\Rc     {\hbox{$R_{\rm c}$}}
\def\Rx     {\hbox{$R_{\rm x}$}}
\def\R     {\hbox{$\cal R$}}
\def\Sl    {\hbox{$S_{\rm c,\lambda}$}}
\def\Sd    {\hbox{$S^{\rm d}_{\rm c,\lambda}$}}
\def\Si    {\hbox{$S^{\rm i}_{\rm c,\lambda}$}}
\def\vc    {\hbox{$v_{\rm cl}$}}
\def\erg   {\hbox{erg\,s$^{-1}$}}
\def\kms   {\hbox{km\,s$^{-1}$}}
\def\MBH   {\hbox{$M_{\bullet\,7}$}}
\def\rpc   {\hbox{$r_{\rm pc}$}}
\def\Mo     {\hbox{$M_{\odot}$}}
\def\Mc    {\hbox{$M_{\rm cl}$}}
\def\Nc     {\hbox{$N_{\rm C}$}}
\def\Ntor   {\hbox{$N_{\rm torus}$}}
\def\NH     {\hbox{$N_{\rm H}$}}
\def\mH    {\hbox{$m_{\rm p}$}}
\def\cc    {\hbox{cm$^{-3}$}}
\def\cs    {\hbox{cm$^{-2}$}}
\def\MaBH  {\hbox{$\dot M_{\rm acc}^{\rm BH}$}}
\def\MaTOR {\hbox{$\dot M_{\rm acc}^{\rm TOR}$}}
\def\Mw    {\hbox{$\dot M_{\rm out}^{\rm TOR}$}}

\def\rpc   {\hbox{$r_{\rm pc}$}}
\def\Myr   {\hbox{$\Mo\,{\rm yr}^{-1}$}}

\begin{document}

\begin{frontmatter}

\title{The Toroidal Obscuration of \\ Active Galactic Nuclei}

\author{Moshe Elitzur}

\address{Department of Physics \& Astronomy,
         University of Kentucky \\
         Lexington, KY 40506-0055,
         USA }

\begin{abstract}
Observations give strong support for the unification scheme of active galactic
nuclei.  The scheme is premised on toroidal obscuration of the central engine
by dusty clouds that are individually very optically thick. These lectures
summarize the torus properties, describe the handling and implications of its
clumpy nature and present speculations about its dynamic origin.

\end{abstract}


\end{frontmatter}

\section{Introduction}

The basic premise of the unification scheme is that every AGN is intrinsically
the same object: an accreting supermassive black hole. This central engine is
surrounded by a dusty toroidal structure so that the observed diversity simply
reflects different viewing angles of an axisymmetric geometry. Since the torus
provides anisotropic obscuration of the center, sources viewed face-on are
recognized as ``type 1'', those observed edge-on are ``type 2''.  The
observational evidence for unification is covered in the lectures by C.
Tadhunter and B. Peterson elsewhere in these proceedings.

A scientific theory must make falsifiable predictions, and AGN unification does
meet this criterion. Unification implies that for every class of type 1 objects
there is a corresponding type 2 class, therefore the theory predicts that type
2 QSO must exist. After many years of searching, the existence of QSO2 has been
established, thanks to the {\em Sloan Digital Sky Survey}, and their
spectro\-polarimetry even reveals the hidden type 1 nuclei at $z$ as large as
0.6. This is a spectacular success of the unification approach. There are not
that many cases in astronomy --- in fact, in all of science --- where a new
type of object has been predicted to exist and then actually discovered.  In
light of this success, it would be hard to question the basic validity of the
unification approach. There is no reason, though, why the obscuring torus
should be the same in every AGN; it is unrealistic to expect AGN's to differ
only in their overall luminosity but be identical in all other aspects.

The torus can be considered an acronym for Toroidal Obscuration Required by
Unification Schemes. From basic considerations, \cite{Krolik88} concluded that
the obscuration is likely to consist of a large number of individually very
optically thick dusty clouds. However, the important roll of clumpiness was not
always fully appreciated, for example, its effect on AGN classification
statistics: type 1 and type 2 viewing is an angle-dependent probability, not an
absolute property. In these lectures I summarize the properties of the
obscuring torus and try to speculate on how it might evolve with the AGN
luminosity, i.e., its accretion rate. Since the main {\em direct} evidence for
the torus comes from its IR emission, I start with a general discussion of dust
emission (sec.\ 2). This is followed by a review of torus phenomenology (sec.\
3), description of the handling of its clumpiness (sec.\ 4) and speculations
about its dynamical origin (sec.\ 5).

\section{Dust--Generalities}

\begin{figure}[ht]
 \centering\leavevmode\includegraphics[width=0.6\hsize,clip]{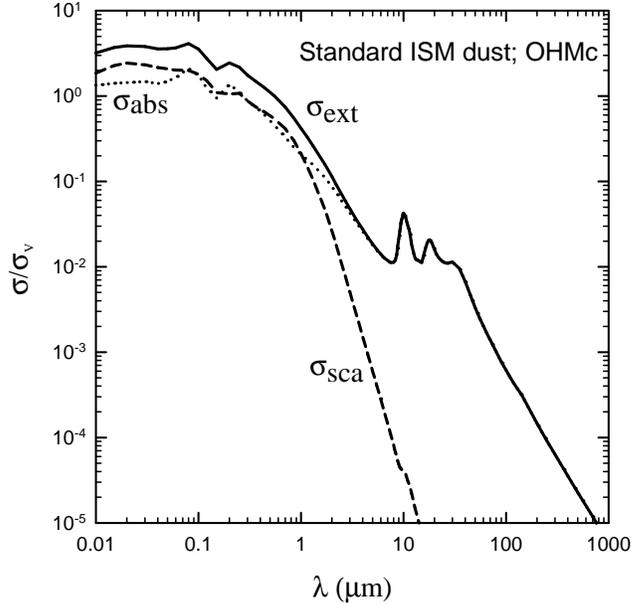}

\caption{Spectral shape of interstellar dust extinction coefficient and its
scattering and absorption components: $\sigma_{\rm ext} = \sigma_{\rm sca} +
\sigma_{\rm abs}$. Standard interstellar grain mix with optical properties from
\cite*{Ossenkopf92} for the silicate component and \cite{Draine03} for the
graphite.} \label{Fig:OHMc}
\end{figure}

Dust grains are mixed with the gas in virtually all interstellar regions. The
grains are small solid particles, built-up through collisions of many atoms
(primarily Si, Mg, Al, C, O) sticking together.  Grain interaction with
radiation is comprised of two components: {\em scattering} and {\em
absorption}; their sum is referred to as {\em extinction}.  Scattering occurs
when radiation is simply reflected off the grain surface, while during
absorption the incoming photon is actually absorbed by the dust grain. The
interaction of dust with radiation follows the classical theory of scattering
by small geometric structures characterized by a dielectric constant.  The
spectral variation of the cross-section $\sigma_\lambda$ is primarily
controlled by the relation between the radiation wavelength $\lambda$ and the
grain size $s$. When $\lambda \ll s$ the grains block the radiation completely
and $\sigma_\lambda$ approaches the constant diffraction limit. When $\lambda
\gg s$, $\sigma_\lambda$ decreases as a power law. Superimposed on this general
behavior are spectral features reflecting internal excitations of the grain
material. Figure \ref{Fig:OHMc} shows the spectral behavior of each of the
three cross sections for the ``standard'' composition of Galactic interstellar
dust: 53\% silicates and 47\% graphite, with a grain size distribution
following a power-law from 0.005\mic\ to 0.25\mic\ \citep*{MRN}. Absorption is
by far the dominant process for interaction of interstellar grains with
radiation in the mid- and far-IR (wavelengths longer than \about\ 2 \mic);
scattering can be safely ignored at these wavelengths.

The dust optical depth at wavelength $\lambda$ is $\tau_\lambda = \int
n_d\sigma_\lambda d\ell$, where $\sigma_\lambda$ is the appropriate
cross-section and $n_d$ is the dust number density. The mean projected area of
dust grains shows the remarkably uniform correlation $n_d\sigma_{\rm V}/n_{\rm
H} \simeq 5\x\E{-22}$ cm$^2$, where $n_{\rm H}$ is the hydrogen density and
$\sigma_{\rm V}$ is the dust extinction cross section at visual (0.55\mic).
This ratio is approximately the same in all Galactic directions
\citep[e.g.,][]{Sparke06}. Dust obscuration with optical depth \tV\ at visual
therefore implies a gas column density \NH\ \about\ 2\x\E{21}\,\tV\ \cs.

\subsection{Dust Temperature}
\label{sec:Tdust}

The interaction of dust grains with radiation determines the state of their
internal excitation; collisional energy exchange with the gas is usually
negligible in the dust energy balance (although it may be important for the
gas).  During scattering the radiation is reflected off the grain surface and
has no effect on the internal energy of the dust particle.  However, when a
grain absorbs an incoming photon it heats up and the energy is shared among the
many internal excitation modes. In steady state, the grains must radiate away
all the energy that they absorb. The balance between absorption and emission
provides a condition of {\em radiative equilibrium}, which determines the grain
temperature $T$: $\int\sigma_{\rm abs\lambda}B_\lambda(T)d\lambda =
\int\sigma_{\rm abs\lambda}J_\lambda d\lambda$, where $B_\lambda$ is the Planck
function and $J_\lambda = \int I_\lambda d\lambda/4\pi$ is the angle-averaged
intensity. The latter can be split into its external and diffuse components
according to $J_\lambda = J_{\rm e\lambda} + J_{\rm diff\lambda}$. When the
external radiation originates from a point source with luminosity $L$ at a
distance $r$, the bolometric magnitude of the external angle-averaged intensity
on the illuminated surface of the dust distribution is $J_e = L/16\pi r^2$.
Denote the Planck-average of the absorption coefficient $\sigma_{\rm
abs\lambda}$ by $\sigma_{\rm aP}$ and its average with the spectral shape of
the external radiation by $\sigma_{\rm ae}$. The dust temperature equation is
then
\eq{\label{eq:Td}
    {\sigma_{\rm aP}(T)\over\sigma_{\rm ae}}\sigma_{\rm B} T^4 =
    {L\over 16\pi r^2}(1 + \delta), \quad \hbox{where} \quad
    \delta = {1\over\sigma_{\rm ae} J_e}
    \int\sigma_{\rm abs\lambda}J_{\rm diff\lambda} d\lambda
}
and where $\sigma_{\rm B}$ is the Stefan-Boltzmann constant. The contribution
of the diffuse radiation, contained in the term $\delta$, couples this relation
to the radiative transfer equation, and the dust temperature cannot be
determined without solving the full dust radiative transfer problem. However,
when the dust optical depth is small then the contribution of the diffuse
radiation is negligible and $\delta \ll 1$. In that case eq.\ \ref{eq:Td} has
an immediate solution, which gives roughly $T \sim r^{-0.4}$. Dust at
temperature $T$ emits predominantly around the peak of the Planck function,
$\lambda \sim 10\mic\,(300\,{\rm K}/T)$. Because of the temperature decline
with distance from the heat source, the wavelength of peak emission from
distance $r$ increases as \about\ $r^{0.4}$. As the dust optical depth
increases, the temperature decline becomes steeper immediately behind the
illuminated surface. However, once the temperature decreases to the point that
the remaining dust is optically thin at wavelengths longer than the peak of the
local Planck function, the optically-thin temperature variation is recovered
\citep[see, e.g.,][]{IE97}.

As the grain size increases, the roughly flat portion of the cross section
evident in figure \ref{Fig:OHMc} extends to longer wavelengths. The result is a
larger $\sigma_{\rm aP}$ and a lower grain temperature. Therefore, individual
grains exposed to a heating radiation will have different temperatures,
decreasing as the grain size increases, and the concept of a single ``dust
temperature'' is meaningless. However, heating of all grains is coupled through
the contribution of the diffuse radiation, and the different temperatures tend
to equilibrate toward a single, common value inside sources with large optical
depths. A commonly used approximation is to replace the grain mixture with a
single-type composite grain whose radiative constants are constructed from the
mix average. This method reproduces adequately full calculations of grain
mixtures, especially when optical depths are large \citep[e.g.,][]{EfRR94,
Wolf03}. The handling of the dust optical properties is exact in this approach,
the only approximation is in replacing the temperatures of the different grain
components with a single average. \cite{Wolf03} finds that the temperatures of
different grains are within $\about \pm10\%$ of the temperature obtained in the
mean grain approximation, and that these deviations disappear altogether when
$\tV > 10$.

\begin{figure}[ht]
 \centering\leavevmode\includegraphics[width=0.7\hsize,clip]{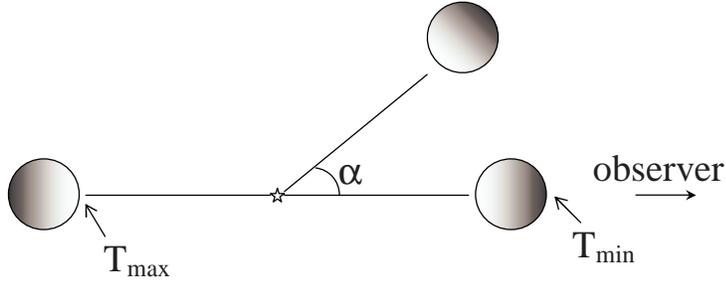}

\caption{The dust temperature varies on the surface of an optically thick cloud
in the radiation field of an external source from $T_{\rm max}$ on the
illuminated face to $T_{\rm min}$ on the dark side. Therefore the cloud's IR
emission is anisotropic. The visible fraction of the illuminated surface area
is determined by the position angle $\alpha$ with respect to the observer
direction.}
 \label{Fig:T clump}
\end{figure}

Dust temperature distributions are profoundly different in clumpy and smooth
environments. In smooth density distributions, dust temperature and distance
from the radiative heating source are uniquely related to each other---given
the distance, the dust temperature (either that of individual grain types or
the common equilibrium value) is known, and vice versa. Clumpy media made up of
discrete clouds that are individually optically thick and have dimensions that
are much smaller than typical inter-cloud distances behave entirely
differently. A cloud heated by a distant source will have a higher temperature
on its illuminated face than on other surface areas (figure \ref{Fig:T clump}).
The result is that in a clumpy medium, different dust temperatures coexist at
the same distance from the radiation source, and the same dust temperature
occurs at different distances --- the dark side of a nearby cloud can be as
warm as the bright side of a farther cloud. This has important implications for
the IR emission from the AGN torus.

Dust grains evaporate when they become too hot. Taking $T_{\rm sub}$ = 1500 K
as a typical sublimation temperature, the shortest wavelength of dust emission
is \about\ 2\mic; emission at shorter wavelengths requires dust hotter than
\about 1500 K, thus it necessarily reflects pure scattering. Inserting the
spectral shape of the nuclear UV/optical radiation from active galaxies as the
heat source in eq.\ \ref{eq:Td} yields the AGN dust sublimation radius
\eq{\label{eq:Rd}
    \Rd \simeq 0.4\left(L\over\E{45}\,{\rm erg\,s^{-1}}\right)^{\!\!1/2}
                  \left(1500\,{\rm K} \over T_{\rm sub}\right)^{\!\!2.6}
                  \ \rm pc
}
\citep{AGN2}. This value of \Rd\ was determined from the temperature on the
illuminated face of an optically thick dusty cloud in the composite grain
approximation. In fact, larger grains are cooler and can survive closer to the
heat source. The sharp boundary defined in eq.\ \ref{eq:Rd} is an
approximation. In reality, the transition between the dusty and dust-free
environments is gradual because individual components of the mix sublimate at
slightly different radii, with the largest grains surviving closest to the AGN.

\subsection{The 10\mic\ Feature}
\label{sec:10mic}

\begin{figure}[ht]
 \centering\leavevmode\includegraphics[width=0.7\hsize,clip]{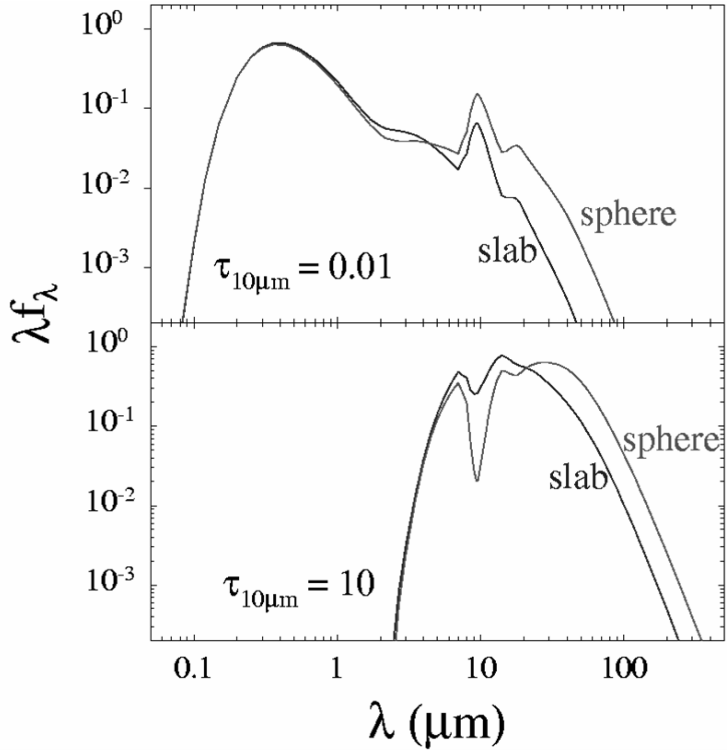}

\caption{Model results for dust with standard interstellar composition
illuminated by black-body radiation with temperature of 10,000 K.  The models
labeled ``sphere'' have dust in a spherical shell with density profile $1/r^2$
around a central point source. In those labeled ``slab'', parallel rays impinge
on the dusty slab. The figure shows the normalized SED ($\int f_\lambda
d\lambda = 1$) of the radiation observed from outside the sphere and the
non-illuminated face of the slab. The dust optical depth across the slab and
along the shell radius is the same in each panel, as marked. \label{Fig:10mic}}
\end{figure}

Superimposed on the generally smooth spectral shape of the dust cross-section
are two prominent features in the mid-infrared (fig.\ \ref{Fig:OHMc}). They are
attributed to amorphous silicate grains that produce strong opacity peaks due
to the Si--O stretching and the O--Si--O bending modes centered around 10 and
18 \mic, respectively. These silicate features, especially the 10\mic\ one, are
visible in the observed spectral energy distribution (SED) of many astronomical
objects, where they appear in either emission or absorption. This variety
arises from radiative transfer effects as demonstrated in figure
\ref{Fig:10mic}, which shows the results of exact model
calculations\footnote{The calculations were performed with the code DUSTY,
which is publicly available at http://www.pa.uky.edu/$\sim$moshe/dusty}.

The feature behavior is affected both by the dust optical thickness and its
geometry. It can be understood with the aid of the simple solution of the
radiative transfer equation for the emission from dust with uniform temperature
$T$ and optical depth $\tau_\lambda$:
\eq{\label{eq:uniformT}
    I_\lambda = B_\lambda(T)\left(1 - e^{-\tau_\lambda}\right)
    \simeq \cases{\tau_\lambda B_\lambda & when $\tau_\lambda \ll 1$ \cr
                                                                    \cr
                  B_\lambda              & when $\tau_\lambda \gg 1$;
                 }
}
when the dust is exposed to external radiation with intensity $I_e$, the
transmission term $I_e\, e^{-\tau_\lambda}$ is added. This result shows that
the emission from optically thin dust ($\tau_\lambda \ll 1$) is proportional to
the dust cross section, therefore the emergent radiation in this case follows
the spectral shape shown in figure \ref{Fig:OHMc}. A superposition of different
dust temperatures preserves the proportionality with $\tau_\lambda$, producing
an emission feature in all optically thin sources. This feature is evident in
the top panel of figure \ref{Fig:10mic}. The optical depth of the solutions
displayed in this panel is rather small (\tV\ = 0.15), and the short
wavelengths ($\lambda$ \la\ 2\mic) emission is simply the input radiation,
which is barely attenuated. Longer wavelengths show the dust emission, and the
difference between the sphere and the slab results is an important reflection
of the different temperature profiles in the two geometries. In a spherical
shell, the dust temperature decreases with radial distance because of the
spatial dilution of the heating radiation (eq.\ \ref{eq:Td}). The large range
of temperatures spanned by the cool dust is reflected in the relatively broad
shoulder of long-wavelength emission in the sphere solution. In contrast, the
slab is exposed to parallel-rays radiation, which does not suffer any spatial
dilution, therefore its temperature is uniform and the dust emission covers a
more limited spectral range.

Optically thin emission increases linearly with the overall optical depth. As
this optical depth approaches unity, self-absorption sets in, higher order
terms in the expansion of $e^{-\tau_\lambda}$ become important and, as is
evident from eq.\ \ref{eq:uniformT}, the emission saturates at the featureless
Planck function: at constant temperature, emission and self-absorption exactly
balance each other, producing the black-body thermodynamic limit. Therefore, a
single-temperature optically thick region can never produce a feature, neither
in emission nor absorption. Furthermore, irrespective of optical depth,
single-temperature dust will never produce an absorption feature; it can only
produce an emission feature when optically thin. The emergence of the
absorption feature visible in the bottom panel of figure \ref{Fig:10mic}
reflects the temperature stratification in actual dusty material: as the
radiation propagates from hot regions toward the observer, it passes through
cooler regions where it suffers absorption that is not balanced by the emission
from these cooler regions. Thus, the temperature structure primarily determines
the strength of the absorption feature at large optical depths. The models
displayed in this panel have \tV\ = 150. The input radiation is fully
extinguished within a short distance from the illuminated face, creating a
large temperature gradient close to the surface. Beyond that absorption layer,
the temperature profile resembles the optically thin case and is controlled by
the geometry---steep decline in the spherical shell and near constancy in the
slab. This difference is reflected not only in the longer wavelength emission
from the spherical shell but also the much greater depth of its 10\mic\
absorption feature.

This analysis shows that a temperature gradient is essential for an absorption
feature; deep absorption requires dust geometry conducive to large gradients.
Consider a cloud illuminated from outside by a radiative source. If the cloud's
dimensions are much smaller than the distance to the source, then the heating
flux is constant across the cloud's volume, just as in the slab model. Without
radiative transfer effects, the dust temperature would be uniform throughout
the cloud. In contrast, arranging the same dust in a geometrically thick shell
around the same heating source produces a large temperature gradient because of
the spatial dilution of the flux with radial distance. We can therefore expect
the absorption feature to have only limited depth in the case of externally
illuminated clouds. A deep feature requires the radiation source to be deeply
embedded in dust that is thick both optically and geometrically. These results
have important implications for the study of dust emission from active galaxies
\citep{Levenson07, Sirocky08}.

\section{Torus Phenomenology}

We start with a discussion of torus observations and the physical properties
that can be inferred directly from them.

\subsection{IR Emission}

\begin{figure}[ht]
 \centering\leavevmode\includegraphics[width=0.7\hsize,clip]{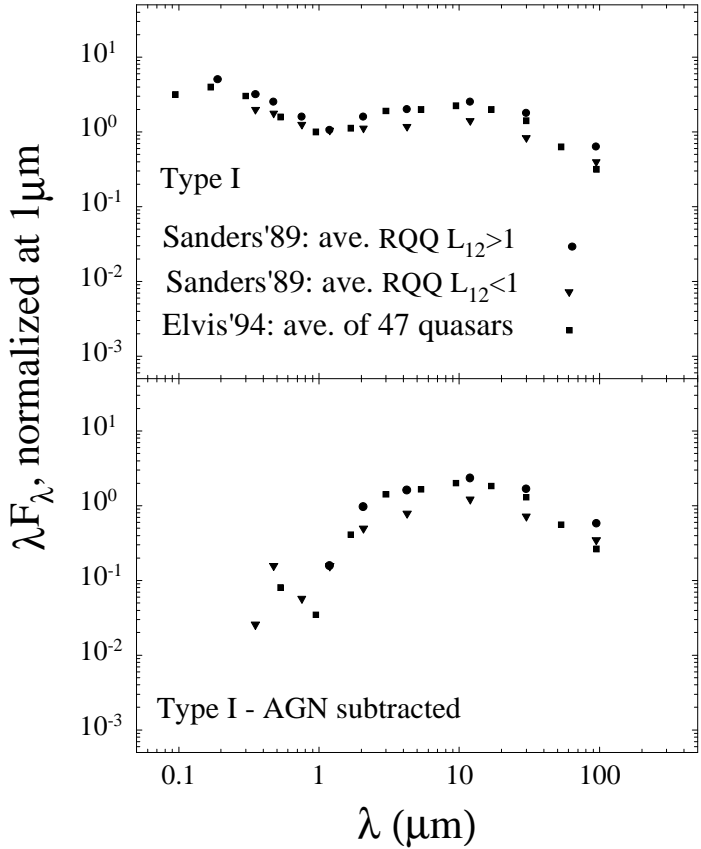}

\caption{SED's of type 1 sources. Top: Average spectra from the indicated
compilations; for the definition of Radio Quiet Quasars (RQQ) see Tadhunter's
lectures. Bottom: The same SED's after subtracting a power law fit through the
short wavelengths ($\la$ 1\mic). These AGN-subtracted SED's are similar to
those observed in type 2 sources. \label{Fig:data1}}
\end{figure}

The primary evidence for the torus comes from spectropolarimetric observations
of type 2 sources, which reveal hidden type 1 emission via reflection off
material situated above the torus opening. While compelling, this evidence is
only indirect in that it involves obscuration, not direct emission by the torus
itself. An obscuring dusty torus should reradiate in the IR the fraction of
nuclear luminosity it absorbs, providing direct evidence for its existence.
Indeed, the continua from most AGNs show significant IR emission. The top panel
of figure \ref{Fig:data1} shows the composite type 1 spectra from a number of
compilations. The optical/UV region shows the power law behavior expected from
a hot disk emission. At $\lambda$ \ga\ 1\mic, the SED shows the bump expected
from dust emission. The bottom panel shows the same data after subtracting a
power law fit through the short wavelengths in a crude attempt to remove the
direct AGN component and mimic the SED from an equatorial viewing of these
sources according to the unification scheme. Indeed, the AGN-subtracted SED's
resemble the observations of type 2 sources. Silicates reveal their presence in
the dust through the 10 \mic\ feature. The feature appears generally in
emission in type 1 sources and in absorption in type 2 sources \citep{Hao07}.

The torus dust emission has been resolved recently in 8--13 \mic\
interferometry with the VLTI. The first torus was detected in NGC 1068 by
\cite{Jaffe04}, the second in Circinus by \cite{Tristram07}. The latter
observations show evidence for the long anticipated clumpy structure. The dust
temperature distributions deduced from these observations indicated close
proximity between hot ($>$ 800 K) and much cooler (\about\ 200--300 K) dust.
Such behavior is puzzling in the context of smooth-density calculations but is
a natural consequence of clumpy models (see sec.\ \ref{sec:Tdust})

\begin{figure}[ht]
 \centering\leavevmode
 \includegraphics[width=0.45\hsize,clip]{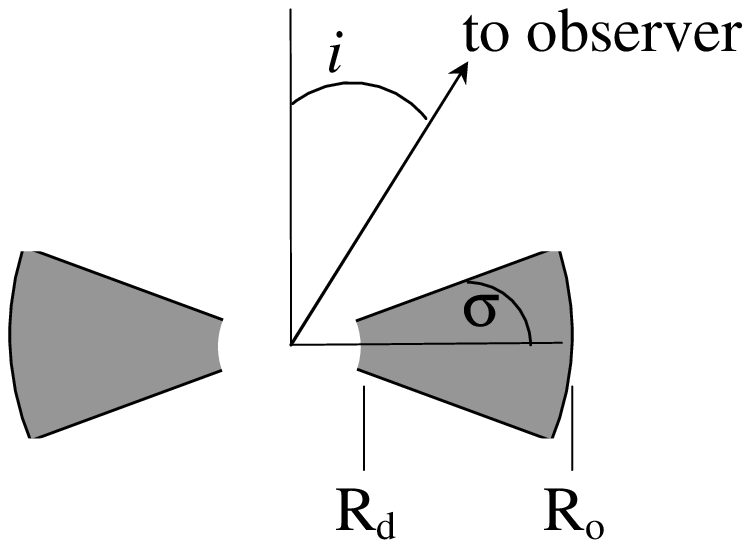} \hfill
 \includegraphics[width=0.45\hsize,clip]{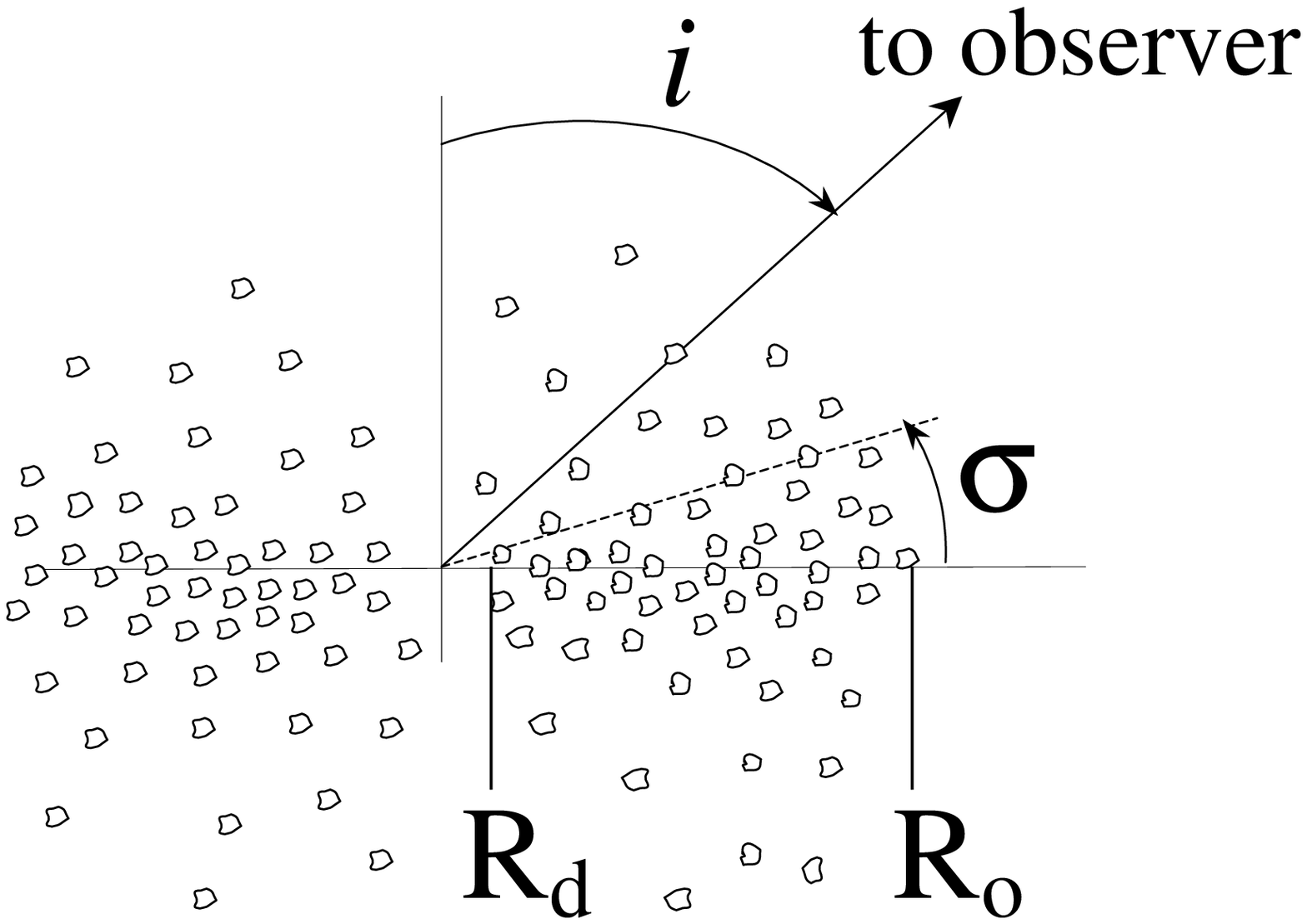}

\caption{AGN classification according to unified schemes. {\em Left}: In a
smooth-density torus, the viewing angle (from the axis) $i = \frac12\pi -
\sigma$ separates between type 1 and type 2 viewing. {\em Right:} In a clumpy,
soft-edge torus, the probability for direct viewing of the AGN decreases away
from the axis, but is always finite.} \label{Fig:Smooth_Clumpy}
\end{figure}

\subsection{Torus Obscuration}
\label{sec:f2}

There are clear indications that the optical depth in the torus equatorial
plane is at least \tV\ \ga\ 10.  If the dust abundance in the torus is similar
to Galactic interstellar regions, the equatorial column density is at least
\NH\ \ga\ 2\x\E{22}~\cs.

The classification of AGN into types 1 and 2 is based on the extent to which
the nuclear region is visible, therefore source statistics can determine the
angular extent of the torus obscuration. In its standard formulation, the
unification approach posits the viewing angle as the sole factor in determining
the AGN type. This is indeed the case for a smooth-density torus that is
optically thick within the angular width $\sigma$ (figure
\ref{Fig:Smooth_Clumpy}, left sketch). All AGN viewed at ${0 \le i < \half\pi -
\sigma}$ then appear as type 1 sources, while viewing at $\half\pi - \sigma \le
i \le \frac12\pi$ gives type 2 appearance. If $f_2$ denotes the fraction of
type 2 sources in the total population, then $f_2 = \sin\sigma$. This relation
holds also when the column density declines smoothly with angle away from the
equatorial plane. The AGN is obscured from directions that have $e^{-\tau} \gg
1$ and visible from those with $e^{-\tau} \ll 1$. Because of the steep
variation of $e^{-\tau}$ with $\tau$, the transition between these two regions
is sharp, occurring around the angle where $\tau = 1$. Denoting this angle
$\sigma$ yields $f_2 = \sin\sigma$ irrespective of the specific angular
profile. Taking account of the torus clumpiness modifies this relation
fundamentally, as is evident from the right sketch in figure
\ref{Fig:Smooth_Clumpy}. We discuss this modification below (sec.\
\ref{sec:unification}).

From statistics of Seyfert galaxies \cite{Schmitt01} find that $f_2 \simeq$
70\%. Employing the standard $f_2 = \sin\sigma$ relation they deduce $\sigma
\simeq$ 45\deg. The issue is currently unsettled because \cite{Hao05b} have
found recently that $f_2$ is only about 50\%, or $\sigma \simeq$ 30\deg. The
fraction $f_2$ of obscured AGN decreases with bolometric luminosity. This has
been verified in a large number of observations that estimate the luminosity
dependence of either $f_2$ or $f_1$ (the fraction of unobscured sources), or
differences between the luminosity functions of type 1 and 2 AGN
\citep[see][]{Hao05b, Simpson05, Maiolino07a}.

\subsection{Torus Size}
\label{sec:size}

Obscuration statistics provide an estimate of the torus angular width $\sigma$.
Denote by $H$ the torus height at its outer radius \Ro, then $H/\Ro =
\tan\sigma$; obscuration does not depend individually on either $H$ or \Ro,
only on their ratio. To determine an actual size one must rely on the torus
emission. In the absence of high-resolution IR observations, early estimates of
the torus size came from theoretical analysis of the SED. \cite{PK92} preformed
the first detailed calculations of dust radiative transfer in a toroidal
geometry, and because of the difficulties in modeling a clumpy medium
approximated the density distribution with a uniform one instead. They
concluded that the torus has an outer radius \Ro\ \about\ 5--10 pc, but later
speculated that this compact structure might be embedded in a much larger, and
more diffuse, torus with \Ro\ \about\ 30--100 pc \citep{PK93}. \cite{Granato94}
extended the smooth-density calculations to more elaborate toroidal geometries.
From comparisons of their model predictions with the observed IR emission at
$\lambda$ \about\ 10--25 \mic\ they concluded that the torus must have an outer
radius \Ro\ \ga\ 300--1000 pc, and that its radial density profile must be
constant; later, \cite{Granato97} settled on hundreds of pc as their estimate
for the torus size. Subsequently, \Ro\ \ga\ 100 pc became common lore.

The advent of high-resolution IR observations brought evidence in support of
Pier \& Krolik's original proposal of compact torus dimensions \citep[for
details, see][]{AGN2}. It can be argued that IR observations only determine the
size of the corresponding emission region and that the actual torus size could
in fact be much larger, but mid-IR flux considerations were the sole reason for
introducing large sizes in the first place. It seems safe to conclude that at
this time there is no compelling evidence that torus clouds beyond \Ro\ \about\
20--30\Rd\ need be considered. From eq.\ \ref{eq:Rd}, a conservative upper
bound on the torus outer radius is then $\Ro < 12\,L_{45}^{1/2}$ pc, where
$L_{45} = L/\E{45}\,\erg$.

\subsection{Torus Orientation and the Host Galaxy}
\label{sec:orientation}

Toroidal obscuration is the cornerstone of the AGN unification scheme. Since
the active nucleus is at the heart of a galaxy, might the obscuration not be
attributed to the host galaxy? In fact, although galactic obscuration can
affect individual sources, the strong orientation-dependent absorption cannot
be generally attributed to the host galaxy because the AGN axis, as traced by
the jet position angle, is randomly oriented with respect to the galactic disk
in Seyfert galaxies \citep{Kinney00} and the nuclear dust disk in radio
galaxies \citep{Schmitt02}. In addition, \cite{Guainazzi01} find that heavily
obscured AGN reside in a galactic environment that is as likely to be
`dust-poor' as `dust-rich'.

NGC 1068, the archetype Seyfert 2 galaxy and one of the most studied active
nuclei, is a revealing example. While the galaxy is oriented roughly face-on,
the torus is edge-on to within \about\,5\deg. The nearly edge-on orientation of
the torus and accretion disk are indicated by the geometry and kinematics of
both water maser \citep{Greenhill97, Gallimore01} and narrow-line emission
\citep{Crenshaw00}. Some other interesting information about the AGN immediate
surroundings in this source comes from molecular line observations.
\cite{Schinnerer00} find from CO velocity dispersions that at $R \simeq$ 70 pc
the height of the molecular cloud distribution is only $H$ \about\ 9--10 pc,
for $H/R$ \about\ 0.15. \cite{Galliano03} model the CO and H$_2$ emission with
a clumpy molecular disk with radius 140 pc and scale height 20 pc, for the same
$H/R$ \about\ 0.15. Thus, although resembling the putative torus, the
distribution of these clouds does not meet the unification scheme requirement
$H/R$ \about\ 1. Evidently, the detected molecular clouds are located in a
thinner disk-like structure outside the torus. \cite{Galliano03} find that this
molecular disk is tilted roughly 15\deg\ from edge-on, much closer to the
orientation of the AGN than of the host galaxy. Imaging polarimetry at 10\mic\
by \cite{Packham07} shed some light on the continuity between the torus and the
host galaxy's nuclear environments. The most recent high resolution adaptive
optics observations of the nuclear region are presented by R. Davies elsewhere
in these proceedings.

\subsection{X-rays}

AGN are strong X-ray emitters, and the X-ray observations give overwhelming
evidence for the orientation-dependent absorption expected from the unification
scheme. The 2--10 keV X-ray continuum from Seyfert 1 nuclei is generally
unattenuated. A small amount of neutral absorption affects some of these
sources, and can be attributed to the host galaxy because it is usually
characterized by typical galactic column densities $N_{\rm H} \la$ \E{21}
cm$^{-2}$ \citep{George98}.\footnote{Spectral features give evidence for ``warm
absorbers'', clouds of ionized gas with $N_{\rm H} \sim$ \E{21}--\E{23}
cm$^{-2}$; see George et al for details.} In contrast, the continuum from
Seyfert 2 galaxies is attenuated by obscuring columns with $N_{\rm H}$ \about\
\E{22}--\E{25} cm$^{-2}$ \citep{Bassani99, Risaliti99}. In accordance with the
unification scheme, the absorption corrected spectra and luminosities of type 2
sources are similar to those of type 1 \citep{Smith96, Turner97a}. From the
correlations they find among various independent absorption indicators,
\cite{Bassani99} conclude that the torus optical thickness along the line of
sight is the parameter most closely correlated with the AGN X-ray properties.
Prominent among these indicators is the K$\alpha$ iron line, the strongest line
in the 4--10 keV X-ray spectrum of AGN. In Seyfert 1 galaxies the line
equivalent width is typically \about 150 eV \citep{Nandra94}, while in Seyfert
2 the equivalent widths are more broadly distributed, ranging from about 100 eV
to 1 keV \citep{Turner97b} and even 5 keV in some Compton-thick AGN
\citep{Levenson02}. The Seyfert 2 equivalent widths reflect processing in a
toroidal structure around the nucleus \citep{Krolik94, Levenson02}.

Yet in spite of the overall correspondence between the optical and X-ray
obscuration, there is a significant number of AGN for which the expected
characteristics are different in the two bands. Although substantial X-ray
absorption is common among type 2 AGN, there are also unabsorbed X-ray sources
that present only narrow emission lines in their optical spectra. Such cases
can be explained with the observational selection effect suggested by
\cite{Severgnini03} and \cite{Silverman05}: in these sources, the optical light
of the host galaxy outshines the AGN continuum and broad lines. This suggestion
was supported by the subsequent studies of \cite{Page06} and \cite{Garcet07}.
In addition, some of the low-luminosity cases may reflect the disappearance of
both torus obscuration and broad line emission (see sec.\ \ref{sec:lowL}
bellow). The opposite case, obscuration only in X-rays, exists too---there are
type 1, broad line AGN with significant X-ray absorption \citep{Perola04,
Eckart06, Garcet07}. Extreme cases include quasars whose optical spectrum shows
little or no dust extinction while their X-ray continuum is heavily affected by
Compton thick absorption \citep{Braito04, Gallagher06}. This cannot be
attributed to observational selection effects.

The differences observed in some sources between X-ray and UV/optical
attenuation arise naturally from the different absorption properties of gas and
dust. Dusty material absorbs continuum radiation both in the UV/optical and
X-rays, therefore the dusty torus provides obscuration of both. But dust-free
gas attenuates just the X-ray continuum, so clouds inside the dust sublimation
radius will provide obscuration only in this band. Conclusive evidence for such
obscuration comes from the short time scales for transit of X-ray absorbing
clouds across the line of sight, which establish the existence of obscuring
clouds inside the dust sublimation radius \citep{Risaliti02}. Extreme cases
involve 4 hour variability \citep{Elvis04} and variations in absorbing column
of more than \E{24} cm$^{-2}$ within two days, indicating Compton thick ({$\NH\
> \E{24} \cs$}) X-ray absorption from a single cloud in the broad-lines region
\citep{Risaliti07}. These observations show that the torus extends inward
beyond the dust sublimation point to some inner radius \Rx\ $<$ \Rd. Clouds at
$\Rx \le r \le \Rd$ partake in X-ray absorption but do not contribute
appreciably to optical obscuration or IR emission because they are dust-free.
Since every cloud that attenuates the optical continuum contributes also to
X-ray obscuration but not the other way round, the X-ray absorbing column
should be at least as large as the UV/optical absorbing column, as observed
\citep{Maccacaro82}. Further, \cite{Maiolino01} find that the X-ray absorbing
column exceeds the reddening column in each member of an AGN sample by a factor
ranging from \about 3 up to \about 100. Although this discrepancy could also
arise from unusual dust-to-gas ratio or dust properties, the observational
evidence for X-ray absorption by clouds in the dust-free inner portion of the
torus implies that in all likelihood, these clouds provide the bulk of the
absorption. This could explain the \cite{Guainazzi05} finding that at least
50\% of Seyfert 2 galaxies are Compton thick.

\subsection{X-ray vs IR}

IR flux measurements collect the emission from the entire torus area on the
plane of the sky, originating from clouds along all rays through the torus. In
contrast, X-ray attenuation is controlled by the clouds along just one
particular ray, the line of sight to the AGN. Therefore, even for the dusty
portion of the obscuring column, the number of X-ray absorbing clouds can
differ substantially from the torus average. Two type 2 sources with similar
cloud properties and the same overall number of clouds following the same
spatial distribution will have an identical IR appearance, yet their X-ray
absorbing columns could still differ significantly. This can be expected to
introduce a large scatter in torus X-ray properties among AGN with similar IR
emission. It may help explain why the SEDs show only moderate variations in the
infrared that are not well correlated with the X-ray absorbing columns
\citep[e.g.,][]{Silva04}.

\section{Clumpiness---Handling and Implications}

The clumpy nature of the environment around the central black-hole is crucial
for understanding various aspects of AGN observations. We start by presenting a
general formalism developed in \cite{NIE02, AGN1} for handling clumpy media.
The original development, described here, applies only to continuum emission
and takes all clouds to be identical, for simplicity.  The formalism was
extended to line emission and to a mixture of cloud properties by
\cite*{Conway05, Conway08}.

\subsection{Clumpy Radiative Transfer}


\begin{figure}[ht]
 \centering\leavevmode\includegraphics[width=0.7\hsize,clip]{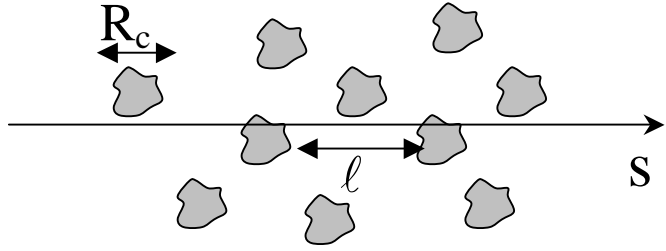}

\caption{A region populated by clouds of size $\Rc$. Along ray $s$, the photon
mean-free-path is $\ell$ and the number of clouds per unit length is \Nc\ =
$\ell^{-1}$. \label{Fig:ClumpyMedium}
}
\end{figure}

\begin{figure}[ht]
 \centering\leavevmode\includegraphics[width=0.7\hsize,clip]{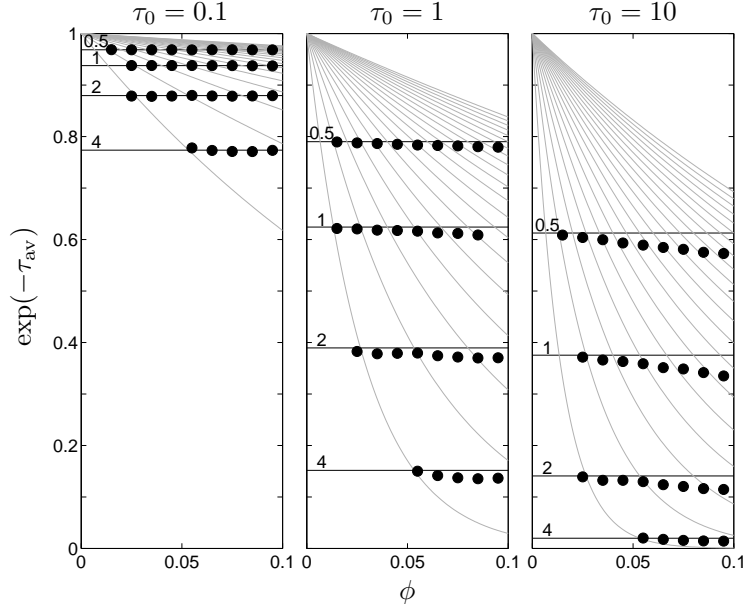}

\caption{Averaged transmission, $\exp(-\tau_{\rm av})$, through a clumpy region
with filling factor $\phi$.  The optical depth of a single cloud is $\tau_0$,
as marked on top each panel; all $\tau_0 \ge 10$ are equivalent. Horizontal
lines show the theoretical small-$\phi$ limit (eq. \ref{eq:Pesc}); they are
labeled by the mean number of clouds along a line through the volume. The
results of Monte Carlo simulations are shown with filled circles whose sizes
are comparable to the numerics error bars. Gray curves trace the loci of
constant cloud sizes. The analytical expression adequately describes clumpy
regions with $\phi$ as large as 10\% \citep{Conway08}. \label{Fig:Rodrigo}
}
\end{figure}

Figure \ref{Fig:ClumpyMedium} sketches a region where the matter is
concentrated in clouds. Individual clouds are characterized by their size \Rc,
the cloud distribution is specified by the number of clouds per unit volume
\nc. It is useful to introduce the number of clouds per unit length \Nc\ =
\nc\Ac\ = $\ell^{-1}$ where \Ac\ is the cloud cross-sectional area and $\ell$
is the photon mean free path for travel between clouds. Denote by \Vc\ the
volume of a single cloud and by $\phi$ the volume filling factor of all clouds,
i.e., the fraction of the overall volume that they occupy. The medium is clumpy
whenever
\eq{\label{eq:clumpy}
 \phi = \nc\Vc  \ll 1\,.
}
In contrast, the matter distribution is smooth, or continuous, when $\phi
\simeq 1$. Since $\Vc \simeq \Ac\Rc$, the clumpiness condition is equivalent to
\eq{\label{eq:clumpy2}
    \phi = \Nc\Rc \ll 1, \qquad \hbox{or}\quad \Rc \ll \ell.
}
The clumpiness criterion is met when the mean free path between clouds greatly
exceeds the cloud size. Under these circumstances, each cloud can be considered
a ``mega-particle'', a point source of intensity \Sl\ and optical depth
$\tau_\lambda$. The intensity at an arbitrary point $s$ along a given path can
then be calculated by applying the formal solution of radiative transfer to the
clumpy medium. The intensity generated in segment $ds'$ around a previous point
$s'$ along the path is $\Sl(s')\Nc(s')ds'$. Denote by $\N(s',s) =
\int_{s'}^s\Nc ds$ the mean number of clouds between $s'$ and $s$ and by
$\Pesc(s',s)$ the probability that the radiation from $s'$ will reach $s$
without absorption by intervening clouds. The number distribution of clouds
between $s'$ and $s$ follows Poisson statistics around the mean $\N(s',s)$, and
\cite{Natta84} show that
\eq{\label{eq:Pesc}
  \Pesc(s',s) = e^{-t_\lambda(s',s)},            \qquad \hbox{where}\quad
  t_\lambda(s',s)  = \N(s',s)( 1 - e^{-\tau_\lambda} ).
}
The intuitive meaning of this result is straightforward in two limiting cases.
When $\tau_\lambda < 1$ we have $t_\lambda(s',s) \simeq \N(s',s)\tau_\lambda$,
which is the overall optical depth between points $s'$ and $s$; that is,
clumpiness is irrelevant when individual clouds are optically thin, and the
region can be handled with the smooth-density approach. It is important to note
that $\N\tau_\lambda$ can be large---the only requirement for this limit is
that each cloud be optically thin. The opposite limit $\tau_\lambda > 1$ gives
$\Pesc(s',s) \simeq e^{-{\cal N}(s',s)}$. Even though each cloud is optically
thick, a photon can still travel between two points  along the path if it
avoids all the clouds in between. With this result, the intensity at $s$
generated by clouds along the given ray is
\eq{\label{eq:IC}
  \IC(s) = \int^s e^{-t_\lambda(s',s)} \Sl(s')\Nc(s') ds'
}
This relation is the exact analog of the formal solution of standard radiative
transfer in continuous media, to which it reverts when the cloud sizes are
decreased to the point that they become microscopic particles; in that case
$\tau \ll 1$ for each particle and $\ell^{-1}$ is the standard absorption
coefficient. The result does not involve $\phi$; only \Nc\ enters. A complete
formalism that would not invoke the assumption $\phi \ll 1$ would lead to a
series expansion in powers of $\phi$, and our derived expression would yield
the zeroth order term in that expansion. In fact, detailed Monte Carlo
simulations show that, to within a few percents, our result describes
adequately clumpy media with $\phi$ as large as 10\%, as shown in figure
\ref{Fig:Rodrigo}. Since the intensity calculations are independent of the
volume filling factor, modeling results do not provide any information on this
quantity, nor do they provide separate information on either \Rc\ or \nc, only
on \Nc. In complete analogy, the radiative transfer problem for smooth density
distributions does not involve separately the size of the dust grains or their
volume density, only the combination $n_{\rm d}\sigma_{\rm d}$, which
determines the absorption coefficient and which is equivalent to \Nc.


An intensity calculation requires the cloud source function \Sl\ at every
location in the clump distribution (eq.\ \ref{eq:IC}). Since the clump emission
is anisotropic, as is evident from figure \ref{Fig:T clump}, \Sl\ depends not
only on location but also on the observer's position angle $\alpha$. It is
convenient to split the population of clumps to those that are directly exposed
to the AGN radiation, with a source function \Sd, and those shadowed by
intervening clouds. The latter are heated only indirectly by the emission from
all other clouds and their source function is denoted \Si. At location
$(r,\beta)$, where $\beta$ is angle from the equatorial plane, the mean number
of clouds to the AGN is $\N(r,\beta) = \int_0^r\!\Nc(r,\beta)\,dr$ and the
probability for unhindered view of the AGN is $p(r,\beta) = e^{-{\cal
N}(r,\beta)}$.  The general expression for the cloud overall source function is
thus
\eq{\label{eq:S}
      \Sl(r,\alpha,\beta) =
      p(r,\beta)\Sd(r,\alpha) + \left[1 - p(r,\beta)\right]\Si(r,\alpha,\beta)
}
The source functions \Sd\ can be calculated from a straightforward solution of
the standard radiative transfer problem once a geometrical shape is assumed for
the individual clouds, a calculation that is independent of the cloud
distribution. The indirect heating of shadowed clouds, on the other hand,
involves the radiation from all other clouds and thus cannot be calculated
without solving the full problem. This is the difficulty presented by diffuse
radiation in standard radiative transfer. Since the diffuse heating is
dominated by the clouds that are directly heated by the AGN, the \Sd\ source
functions can serve as the starting point for an iterative solution of the full
problem: From the emission of the directly heated clouds devise a first
approximation for the diffuse radiation field. Next, place clouds in this
radiation field and calculate their emission to derive a first approximation
for the source functions of indirectly illuminated clouds and, from  eq.\
\ref{eq:S}, the composite source function at every location. In successive
iterations, add to the AGN direct field the cloud radiation calculated from
eq.\ \ref{eq:IC}, and repeat the process until convergence. \cite{NIE02, AGN1}
performed only the first step of this procedure, an approximation that leaves
out the feedback effect of the clouds on their own radiative heating. Carrying
out higher orders of the iteration scheme is an open challenge.

\begin{figure}[ht]
 \centering\leavevmode\includegraphics[width=0.7\hsize,clip]{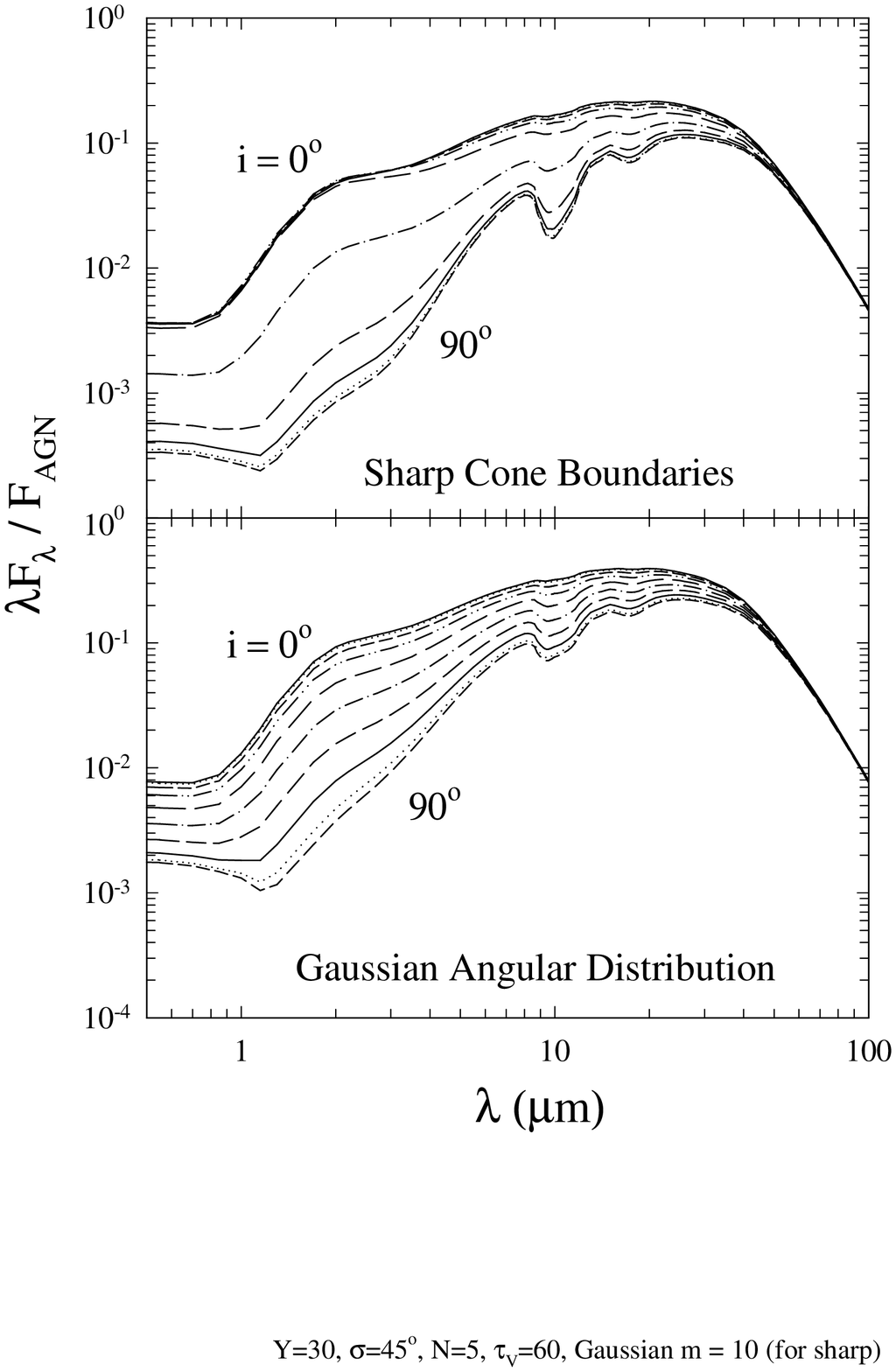}

\caption{Model spectra for dust IR radiation of a clumpy torus. In the bottom
panel the cloud distribution follows eq. \ref{eq:Nc} with \No\ = 5, $\sigma$ =
45\deg\ and $q = 1$ in the radial range $\Rd \le r \le 30\Rd$. Each cloud has
an optical depth \tV\ = 60. Models in the top panel are the same, except that
the angular distribution is uniform within $|\beta| \le \sigma$. Different
curves show viewing angles that vary in 10\deg\ steps from pole-on ($i$ =
0\deg) to edge-on ($i$ = 90\deg). Fluxes are scaled with $\FAGN = L/4\pi D^2$.
\citep{AGN2}} \label{Fig:geometry}
\end{figure}

\subsection{Clumpy Torus Emission}
\label{sec:clumpyIR}

The only distribution required for calculating the intensity from eq.\
\ref{eq:IC} is \Nc, the number of clouds per unit length. The torus axial
symmetry implies that \Nc\ depends only on the distance $r$ from the center and
the angle $\beta$ from the equatorial plane (the complementary of the standard
polar angle). Denote by \No\ the mean of the total number of clouds along
radial rays in the equatorial plane. We parametrize the angular profile of the
cloud distribution as a Gaussian with width parameter $\sigma$ and the radial
distribution as a declining power law with index $q$, so that
\eq{\label{eq:Nc}
    \Nc(r,\beta) = {C\over\Rd}\ \No e^{-\beta^2/\sigma^2}
                   \left({\Rd\over r}\right)^{\!\! q}
}
where $C = (\int_1^Y dy/y^q)^{-1}$ is a dimensionless constant, ensuring the
normalization $\No = \int\Nc(r,\beta = 0) dr$. The Gaussian can be taken as
representative of soft-edge angular distributions, illustrated on the right in
figure \ref{Fig:Smooth_Clumpy}. Sharp-edge angular distributions can be
parametrized with a step function with cutoff at $\beta = \sigma$. Figure
\ref{Fig:geometry} shows the results of model calculations for these two
geometries with the same set of parameters. The sharp-edge geometry produces a
bimodal distribution of spectral shapes, with little dependence on viewing
angle other than the abrupt change that occurs between the torus opening and
the obscured region. This SED dichotomy conflicts with observations
\citep{Almudena03}. In contrast, the soft-edge Gaussian distribution produces a
larger variety in model spectral shapes, with a smooth, continuous dependence
on $i$, in agreement with the findings of \cite{Almudena03}. It is gratifying
that the more plausible soft-edge distribution also produces better agreement
with observations.

\begin{figure}
 \centering\leavevmode\includegraphics[width=0.8\hsize,clip]{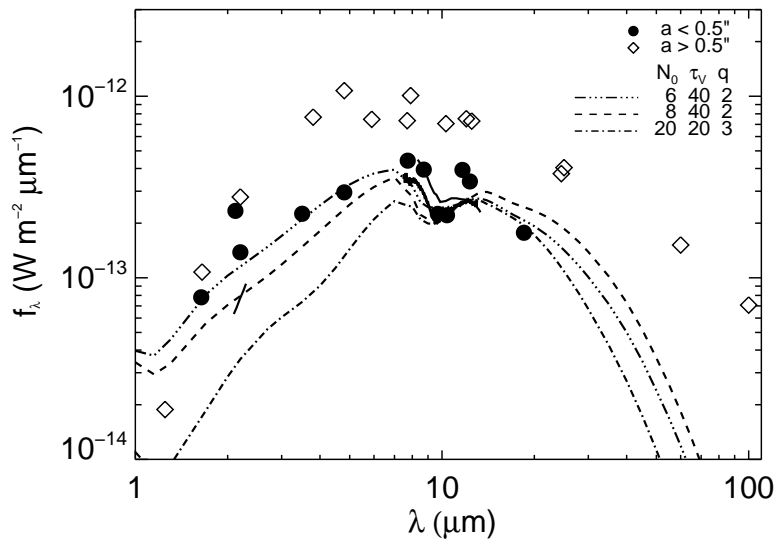}

\caption{Fits of clumpy torus models to small-aperture (0.4$''$) Gemini 8--13
\mic\ spectrum of NGC~1068, shown with heavy solid line. Symbols and thin solid
line between 8--13 \mic\ mark all other data, which were not utilized in the
fits. Filled symbols denote measurements with aperture $a < 0.5''$, open
symbols $a > 0.5''$. Model spectra are for equatorial viewing and various
parameters \citep[for details see][]{Mason06}.} \label{Fig:NGC1068}
\end{figure}

Clumpy torus modeling was employed in the analysis of spatially-resolved,
near-diffraction-limited 10~\mic\ spectra of the nucleus of NGC~1068, obtained
by \cite{Mason06} with Michelle, the mid-IR imager and spectrometer on the
Gemini North telescope. Mason et al first compared the models with the 8--13
\mic\ Michelle spectrum of the central 0.4$''$ core to determine the best-fit
parameters. The model predictions outside this spectral range were then
compared with all other observations, and figure \ref{Fig:NGC1068} shows the
results of this comparison. As noted above, the torus in this AGN is nearly
edge-on (sec.\ \ref{sec:orientation}), and this is the first model to reproduce
the observed near-IR flux with such orientation; in contrast, smooth-density
models were forced to assume, contrary to observations, viewing from
22\deg--25\deg\ above the equatorial plane in order to bring into view the warm
face of the torus backside \citep{Granato97, Gratadour03, Fritz06}. It is
noteworthy that all flux measurements with apertures $< 0.5''$ are in good
agreement with the model results. However, the flux collected with larger
apertures greatly exceeds the model predictions at wavelengths longer than
\about 4\mic. This discrepancy can be attributed to IR emission from nearby
dust outside the torus---Mason et al show that the torus contributes less than
30\% of the 10 \mic\ flux collected with apertures $\ge 1''$. The bulk of the
large-aperture flux comes at these wavelengths from dust in the ionization
cones\footnote{For discussion of ionization cones see lectures by Netzer,
Peterson and Tadhunter.}; while less bright than the torus dust, it occupies a
much larger volume. On the other hand, the torus dominates the emission at
short wavelengths; at 2 \mic, more than 80\% of the flux measured with
apertures $\ge 1''$ comes from the torus even though its image size is less
than 0.04$''$ \citep{Weigelt04}. This highlights a difficult problem that
afflicts all IR studies of AGN. The torus emission can be expected to dominate
the observed flux at near IR because such emission requires hot dust that
exists only close to the center. But longer wavelengths originate from cooler
dust, and the torus contribution can be overwhelmed by the surrounding regions.
There are no easy solutions to this problem.

All in all, clumpy torus models seem to produce SED's that are in reasonable
overall agreement with observations for the following range of parameters
\citep{NIE02, AGN1, AGN2}:
\begin{itemize}
\item
Number of clouds, on average, along radial equatorial rays \No\ = 5--15

\item
Gaussian angular distribution with width $\sigma$ = 30\deg--50\deg

\item
Index of power law radial distribution $q$ = 1--2

\item
Optical depth, at visual, of each cloud \tV\ = 30--100

\item
The torus extends from dust sublimation at \Rd\ = 0.4$L_{45}^{1/2}$ pc to
an outer radius \Ro\ $>$ 5\Rd

\end{itemize}

\begin{figure}[ht]
 \centering \leavevmode
 \includegraphics[width=0.8\hsize,clip]{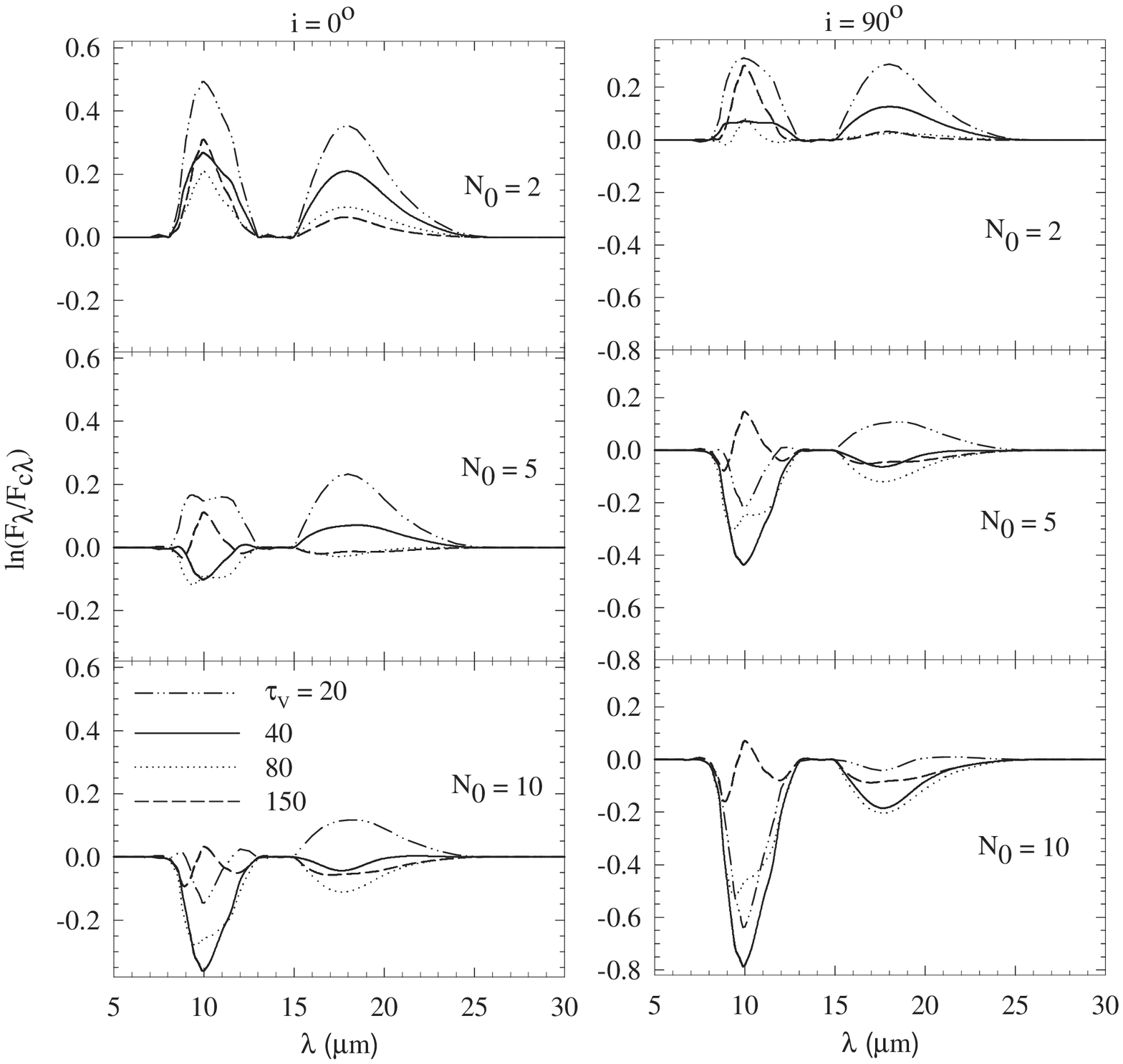}

\caption{Spectral shape of the silicate 10 and 18 \mic\ features: $F_\lambda$
is the torus emission in the 5--30 \mic\ region and \Fcont\ is the smooth
underlying continuum obtained by a spline connecting the feature-free segments
in this spectral region. All models have $q$ = 2, \Ro\ = 30\Rd, $\sigma$ =
45\deg, and \No\ as marked in each panel. Curves correspond to different \tV,
as labeled. Panels on the left correspond to pole-on viewing, on the right to
edge-on; note the different scales of the vertical axes in the two cases.
\citep{AGN2}
} \label{Fig:silfeature}
\end{figure}

The silicate features display in clumpy models a complex behavior that cannot
be reproduced with smooth-density distributions. Figure \ref{Fig:silfeature}
shows some model results for the spectral region containing the features. The
depth of the absorption features is never large in spite of the huge optical
depths of some of the displayed models---a torus with \tV\ = 150 and \No\ = 10
has an equatorial optical depth of 1500 at visual. The relatively shallow
absorption reflects the small temperature gradient across an externally
illuminated cloud, similar to the slab case (see sec.\ \ref{sec:10mic}, fig.\
\ref{Fig:10mic}). Indeed, a striking characteristic of all AGN spectra is the
absence of any deep 10 \mic\ absorption features in contrast with Ultraluminous
IR Galaxies (ULIRGs) where the features reach extreme depths \citep{Hao07}.
This different behavior in ULIRGs can be attributed to deep embedding in a dust
distribution that is smooth, rather than clumpy \citetext{\citealt{Levenson07};
see also \citealt{Spoon07}, \citealt{Sirocky08}}. A most peculiar result is the
emergence of the 10\mic\ feature in \emph{emission} at equatorial viewing when
\tV\ is large for all \No, and when \No\ is small for all \tV. This could
explain intriguing {\em Spitzer} observations of seven high-luminosity type 2
QSOs by \cite{Sturm06}. Although the individual spectra appear featureless, the
sample average spectrum shows the 10 mic\ feature in emission. Therefore, if
this finding is verified it could indicate that the optical depths of torus
clouds perhaps are larger in QSOs than in Seyfert galaxies or that the cloud
number \No\ decreases as the luminosity increases.

While the parameters listed above lead to SED's compatible with observations,
some additional considerations can further restrict the acceptable range. For
example, in spite of the huge difference in AGN obscuration between type 1 and
2 sources, comparisons of their IR emission seem to indicate a surprisingly low
amount of anisotropy \citep{Lutz04, Buchanan06, Horst06}. Models with $q$ = 2
tend to produce less variation with viewing-angle than those with $q$ = 1.
Therefore, observations that will place reliable tight constraints on the IR
emission anisotropy may indicate that $q$ = 2 provides a more appropriate
description of the radial distribution than $q$ = 1. Finally, all model
calculations were performed with standard Galactic ISM dust, which seems to
provide satisfactory results. Current data do not provide any compelling reason
for drastic changes in the dust composition.

\subsection{Clumpy Unification}
\label{sec:unification}

Because of clumpiness, the difference between types 1 and 2 is not truly an
issue of orientation but of probability for direct view of the AGN (figure
\ref{Fig:Smooth_Clumpy}, right sketch); {\em AGN type is a viewing-dependent
probability.} Since that probability is always finite, type 1 sources can be
detected from what are typically considered type 2 orientations, even through
the torus equatorial plane: if \No\ = 5, for example, the probability for that
is $e^{-5} = 1/148$ on average. This might offer an explanation for the few
Seyfert galaxies reported by \cite{Almudena03} to show type 1 optical line
spectra together with 0.4--16 \mic\ SED that resemble type 2. Conversely, if a
cloud happens to obscure the AGN from an observer, that object would be
classified as type 2 irrespective of the viewing angle. In cases of such single
cloud obscuration, on occasion the cloud may move out of the line-of-sight,
creating a clear path to the nucleus and a transition to type 1 spectrum. Such
transitions between type 1 and type 2 line spectra have been observed in a few
sources \citep[see][and references therein]{Aret99}. It is worth while to
conduct monitoring observations in an attempt to detect additional such
transitions. The most promising candidates would be obscured systems with
relatively small X-ray obscuring columns, small torus sizes (low AGN
luminosity; see eq.\ \ref{eq:Rd}) and large black-hole masses \citep{AGN2}.

\begin{figure}
 \centering\leavevmode\includegraphics[width=0.7\hsize,clip]{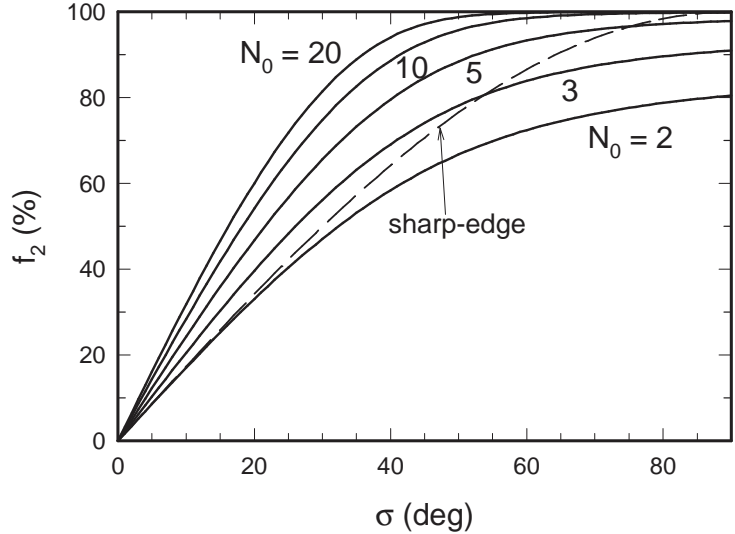}

\caption{AGN statistics: The fraction $f_2$ of obscured sources as a function
of the torus width parameter $\sigma$. In torii with either smooth density
distribution or a sharp-edge clumpy one, this fraction is determined uniquely
by $\sigma$, and is shown with the dashed line. In contrast, in a clumpy torus
with Gaussian angular distribution (eq.\ \ref{eq:Nc}), $f_2$ depends on both
$\sigma$ and the cloud number \No, which is marked on the various solid lines
\citep{AGN2}.
} \label{Fig:f2}
\end{figure}

A sharp-edge clumpy torus has $f_2 = (1 - e^{-{\cal N}_0})\sin\sigma$, which is
practically indistinguishable from the smooth-density relation when $\No$
exceeds \about\ 3--4 (see sec.\ \ref{sec:f2}). However, the situation changes
fundamentally for soft-edge clumpy distributions because at every viewing
angle, the probability of obscuration increases with the number of clouds. As
is evident from figure \ref{Fig:f2}, the Gaussian distribution produces a
strong dependence on \No\ and significant differences from the sharp-edge case.
Since the sharp-edge angular distribution is ruled out by observations, {\em
the fraction of obscured sources depends not only on the torus angular width
but also on the average number of clouds along radial rays}. While the fraction
$f_2 = 70\%$ requires $\sigma = 45\deg$ in the smooth-density case, it implies
$\sigma = 33\deg$ in a Gaussian clumpy torus with \No\ = 5 clouds.

Observations show that $f_2$ decreases with $L$. This has been interpreted as
support for the ``receding torus'' model, in which $\sigma$ decreases with $L$
(Simpson 2005 and references therein). However, all the quantitative analyses
performed thus far for the $L$-dependence of $f_2$ were based on the sharp-edge
expression $f_2 = \sin\sigma$. Removing this assumption affects profoundly the
foundation of the receding torus model because the dependence on the number of
clouds necessitates analysis with two free parameters, therefore $\sigma$
cannot be determined without \No. A decrease of \No\ with $L$ at constant
$\sigma$ will also produce a decrease in $f_2$, the same effect as a decrease
of $\sigma$ (figure \ref{Fig:f2}). An observed trend of $f_2$ with $L$ may
arise from a dependence on either $\sigma$ or \No\ or both. There is no obvious
a-priori means for deciding between the various possibilities.

\subsection{Torus Mass; Total Number of Clouds}

The total mass in torus clouds can be written as $\Mtor =
\mH\NH\int\!\!\Nc(r,\beta)\,dV$, where \mH\ is the proton mass and \NH\ is the
column density of a single cloud; note that \Mtor\ does not involve the volume
filling factor. Denote $Y = \Ro/\Rd$ the torus radial thickness. With the cloud
distribution from eq.\ \ref{eq:Nc} and taking for simplicity a sharp-edge
angular distribution, so that the integration is analytic, the torus mass is
$\Mtor = 4\pi\mH\sin\sigma\Ntor\Rd^2 Y\, I_q(Y)$, where $I_q$ = 1, $Y/(2\ln Y)$
and $\frac13 Y$ for $q$ = 2, 1 and 0, respectively. Taking \Rd\ from eq.\
\ref{eq:Rd}, the mass ratio of the torus and the central black hole is
\eq{\label{eq:Mtor}
  {\Mtor\over M_\bullet} = 2\x\E{-4} {L\over\LEdd} \sin\sigma\Ntor_{,23}\,
                           Y\,I_q
}
where \LEdd\ is the Eddington luminosity and $\Ntor_{,23}$ is the equatorial
column density in \E{23} cm$^{-2}$. Since the Eddington ratio $L/\LEdd$ is
always $<$ 1 and the radial thickness $Y$ is in all likelihood no larger than
\about 20--30 (sec.\ \ref{sec:size}), the torus mass is always negligible in
comparison with $M_\bullet$ when $q$ = 2. If the radial cloud distribution is
flatter, eq.\ \ref{eq:Mtor} may constrain the torus properties to keep its mass
below that of the black-hole.

The total number of torus clouds is $n_{\rm tot} = \int\nc dV =
\int(\Nc/\Ac)dV$. This is the only quantity of interest that depends explicitly
on the cloud size. Equivalently, \Rc\ can be replaced by the volume filling
factor $\phi$, since inserting eq.\ \ref{eq:Nc} into eq.\ \ref{eq:clumpy2}
yields \Rc\ = $C\phi$\Rd/\No\ at the torus inner edge. If $\phi$ is constant
throughout the torus then $\ntot \simeq \N_0^3/\phi^2$ for the $1/r^2$
distribution, independent of the torus radial thickness $Y$. For example, if
the volume filling factor is 10\%, in order to encounter \No\ = 5--10 clouds
along each radial equatorial ray, the torus must contain \ntot\ $\simeq$
\E4--\E5 clouds,

\section{What is the Torus?}

In the ubiquitous sketch by \cite{Urry95}, shown numerous times also in this
summer school, the AGN central region, comprised of the super\-massive black
hole (SBH), its accretion disk and the broad-line emitting clouds, is
surrounded by a large doughnut-like structure---the torus. This hydrostatic
object is a separate entity, presumably populated by molecular clouds accreted
from the galaxy. Gravity controls the orbital motions of the clouds, but the
origin of vertical motions capable of sustaining the ``doughnut'' as a
hydrostatic structure whose height is comparable to its radius was recognized
as a problem since the first theoretical study by \cite{Krolik88}. An entirely
different scenario, hydrodynamic rather than hydrostatic, was proposed by
\cite{Emmering92}. It involves the outflow of clouds embedded in a
hydromagnetic disk wind, avoiding the vertical support problem. The clouds are
accelerated by the system rotation along magnetic field lines anchored in the
disk in a manner first described in the \cite{Blandford_Payne} seminal work. In
this approach the torus is merely a region in the wind which happens to provide
the required toroidal obscuration because the clouds there are dusty and
optically thick. That is, instead of a hydrostatic ``doughnut'', the torus is
just one region in a clumpy wind coming off the black-hole accretion disk. The
wind paradigm was strongly advocated by A. K\"onigl and his associates
\citep{Konigl94, Kartje96, Kartje99, Everett05} and by \cite{Bottorff97,
Bottorff00} but did not gain wide acceptance until recently because of the
general perception that the torus must extend to more than \about 100 pc, a
distance too large to support the required disk wind.

Recent observations suggest that the disk wind scenario might offer the better
paradigm for the AGN torus \citep{Elitzur_Shlosman}. High-resolution IR
observations indicate that the torus size might actually be no more than a few
pc (sec.\ \ref{sec:size}). This compact size places the torus clouds well
inside the black hole sphere of influence, where the SBH gravity dominates over
the galactic bulge. The outer boundary of this region is where the SBH and the
bulge induce gravitational rotations with equal angular velocities, i.e., the
radius within which the bulge mass is equal to that of the black hole. Consider
a spherical bulge that induces a linearly rising rotation curve with $\Omega
\sim 1$\,\kms\,pc$^{-1}$, as is typical of AGN host galaxies \citep{Sofue99},
and an SBH with a mass of \MBH\x\E7\,\Mo\ at its center. The SBH will dominate
the gravitational motions within a radius 35\,pc\x$(\MBH/\Omega_1^2)^{1/3}$,
where $\Omega_1 = \Omega/(1$\,\kms\,pc$^{-1})$. Since the torus is well within
this sphere of influence, its dynamic origin is determined in all likelihood
not by the host galaxy but by the central engine and its accretion disk. There
is mounting evidence for cloud outflow in these regions with the geometry and
kinematics of disk winds \citep[e.g.,][]{Elvis_winds, Gallagher07}, in accord
with the outflow paradigm.

Two different types of observations show that the torus is a smooth
continuation of the broad lines region (BLR), not a separate entity. IR
reverberation observations by \cite{Suganuma06} measure the time lag of the
dust radiative response to temporal variations of the AGN luminosity,
determining the torus innermost radius. Their results show that this radius
scales with luminosity as $L^{1/2}$ and is uncorrelated with the black hole
mass, demonstrating that the torus inner boundary is controlled by dust
sublimation (eq.\ \ref{eq:Rd}), not by dynamical processes. Moreover, in each
AGN for which both data exist, the IR time lag is the upper bound on all time
lags measured in the broad lines, a relation verified over a range of \E6\ in
luminosity. This finding shows that the BLR extends outward all the way to the
inner boundary of the dusty torus, validating the \cite{Netzer_Laor} proposal
that the BLR size is bounded by dust sublimation. The other evidence is the
finding by \cite{Risaliti02} that the X-ray absorbing columns in Seyfert 2
galaxies display time variations caused by cloud transit across the line of
sight. Most variations come from clouds that are dust free because of their
proximity ($<$ 0.1 pc) to the AGN, but some involve dusty clouds at a few pc.
Other than the different time scales for variability, there is no discernible
difference between the dust-free and dusty X-ray absorbing clouds, nor are
there any gaps in the distribution. These observations show that the X-ray
absorption, broad line emission and dust obscuration and reprocessing are
produced by a single, continuous distribution of clouds. The different
radiative signatures merely reflect the change in cloud composition across the
dust sublimation radius \Rd. The inner clouds are dust free. Their gas is
directly exposed to the AGN ionizing continuum, therefore it is atomic and
ionized, producing the broad emission lines. The outer clouds are dusty,
therefore their gas is shielded from the ionizing radiation, and the atomic
line emission is quenched. Instead, these clouds are molecular and dusty,
obscuring the optical/UV emission from the inner regions and emitting IR. Thus
the BLR occupies $r < \Rd$ while the torus is simply the $r > \Rd$ region. Both
regions absorb X-rays, but because most of the clouds along each radial ray
reside in its BLR segment, that is where the bulk of the X-ray obscuration is
produced. Since the unification torus is just the outer portion of the cloud
distribution and not an independent structure, it is appropriate to rename it
the TOR for Toroidal Obscuration Region. The close proximity of BLR and TOR
clouds should result in cases of partial obscuration, possibly leading to
observational constraints on cloud sizes.

The merger of the ionized and the dusty clouds into a single population fits
naturally into the TOR outflow paradigm. The AGN accretion disk appears to be
fed by a midplane influx of cold, clumpy material from the main body of the
galaxy \citep[][and references therein]{Shlosman90}. Approaching the center,
conditions for developing hydromagnetically- or radiatively-driven winds above
this equatorial inflow become more favorable. The disk-wind rotating geometry
provides a natural channel for angular momentum outflow from the disk
\citep{Blandford_Payne} and is found on many spatial scales, from protostars to
AGN. The composition along each streamline reflects the origin of the outflow
material at the disk surface. The disk outer regions are dusty and molecular,
as observed in water masers in some edge-on cases \citep{Greenhill05}, and
clouds uplifted from these outer regions feed the TOR. Such clouds have been
detected in water maser observations of Circinus \citep{Greenhill03} and NGC
3079 \citep{Kondratko05}.  The Circinus Seyfert 2 core provides the best
glimpse of the AGN dusty/molecular component. Water masers trace both a
Keplerian disk and a disk outflow \citep{Greenhill03}. Dust emission at
8--13\mic\ shows a disk embedded in a slightly cooler and larger, geometrically
thick torus \citep{Tristram07}. The dusty disk coincides with the maser disk in
both orientation and size. The outflow masers trace only parts of the torus.
The lack of full coverage can be attributed to the selectivity of maser
operation---strong emission requires both pump action to invert the maser
molecules in individual clouds and coincidence along the line of sight in both
position and velocity of two maser clouds \citep{Kartje99}. In NGC~3079, four
maser features were found significantly out of the plane of the maser-traced
disk yet their line-of-sight velocities reflect the velocity of the most
proximate side of the disk. \cite{Kondratko05} note that this can be explained
if, as proposed by \cite{Kartje99}, maser clouds rise to high latitudes above
the rotating structure while still maintaining, to some degree, the rotational
velocity imprinted by the parent disk. Because the detected maser emission
involves cloud-cloud amplification that requires precise alignment in both
position and velocity along the line-of-sight, the discovery of four
high-latitude maser features implies the existence of many more such clouds
partaking in the outflow in this source.  Moving  inward from the
dusty/molecular regions, at some smaller radius the dust is destroyed and the
disk composition switches to atomic and ionized, producing a double-peak
signature in some emission line profiles \citep{Eracleous04}. The outflow from
the atomic/ionized inner region feeds the BLR and produces many atomic line
signatures, including evidence for the disk wind geometry \citep{Hall03}.

The outflow paradigm for the obscuring torus provides what can be called a
``Grand Unification Theory'' for AGN. This picture requires only the accretion
disk and the clumpy outflow it generates---there is no need to invoke any
additional, separate structure such as a hydrostatic ``doughnut''.  Cloud
radial distance from the AGN center and vertical height above the accretion
disk explain the rich variety of observed radiative phenomena. In both the
inner and outer outflow regions, as the clouds rise and move away from the disk
they expand and lose their column density, limiting the vertical scope of X-ray
absorption, broad line emission and dust obscuration and emission. The result
is a toroidal geometry for both the BLR and the TOR. With further rise and
expansion, the density decreases so the ionization parameter increases, turning
the clouds into members of the warm absorber population (see lectures by H.\
Netzer elsewhere in these proceedings). Similar considerations can be invoked
to explain the broad absorption lines observed in some quasars
\citep[BAL/QSO;][]{Gallagher07}. We proceed now to discuss some consequences of
the TOR outflow scenario.

\subsection{TOR Cloud Properties}
\label{sec:TOR clouds}

The only property of individual clouds constrained from the IR modeling is
their optical depth. It should lie in the range \tV\ \about\ 30--100  (sec.
\ref{sec:clumpyIR}), i.e., the cloud column density is \NH\ \about\
\E{22}--\E{23} \cs\ assuming standard dust-to-gas ratio. Clouds uplifted into
the wind expand while moving away from the disk and their column density
decreases. A cloud starting with \NH\ \about\ \E{23} cm$^{-2}$ ceases to
partake in obscuration when its column is reduced by \about\ 100 so that its
\tV\ drops below unity. Clouds starting with a smaller column will rise to a
smaller height before losing their obscuration and leaving the TOR population.
The obscuration properties constrain only the cloud column density, the product
of the cloud's size and density. The latter can be constrained separately when
we note that the cloud must be able to withstand the black-hole tidal shearing
effect. Consider the tidal torque of the central SBH at a distance \rpc\ in pc,
where the Keplerian period is $t_{\rm K} = 3\x\E4 M_{\bullet 7}^{-1/2} r_{\rm
pc}^{3/2}~{\rm yr}$. To prevent a cloud with density $\rho = \mH n$ from
shearing, it must be at least partially confined by its own gravity and/or the
ambient magnetic field $B$. In the former case, the characteristic Jeans
timescale is $t_{\rm J} \sim (G\rho)^{-1/2} = 3\x\E4 n_7^{-1/2}~{\rm yr}$,
where $n_7 = n/(\E7\,\cc)$. Therefore, resistance to tidal shearing ($t_{\rm J}
< t_{\rm K}$) sets a lower limit on the density of the cloud, leading to upper
limits on its size \Rc\ and mass \Mc:
\eq{\label{eq:cloud}
 n_7 > {\MBH\over r_{\rm pc}^3}, \quad
 \Rc\ \la\ \E{16}{N_{\rm H,23}r_{\rm pc}^3 \over \MBH}\, {\rm cm}, \quad
 \Mc\ \la\ 7\x\E{-3}N_{\rm H,23}\Rc_{16}^2\,\Mo
}
where $\Rc_{16} = \Rc/(\E{16}\,\rm cm)$ and $N_{\rm H,23} = \NH/(\E{23}\,\cs)$.
The resistance to tidal shearing does not guarantee confinement against the
dispersive force of internal pressure. Self-gravity cannot confine these clouds
because their low masses provide gravitational confinement only against
internal motions with velocities \la\ 0.1 \kms. A corollary is that these
clouds cannot collapse gravitationally to form stars. While self-gravity cannot
hold a cloud together against dispersal, an external magnetic field $B \sim
1.5\,\sigma_5n_7^{1/2}$ mG would suffice if the internal velocity dispersion is
$\sigma_5\x$1\,\kms. Clouds with these very same properties can explain the
masers detected in Circinus and NGC~3079, adding support to the suggestion that
the outflow water masers are yet another manifestation of the dusty, molecular
clouds that make up the torus region of the disk-wind. Densities of H$_2$O
masers are frequently quoted as \about\ \E8--\E9\ \cc, but these are the
optimal densities to produce the highest possible brightness from a single
cloud. Because of the scaling properties of H$_2$O pumping \citep{EHM89}, maser
clouds with $n$ \about\ \E7\ \cc\ and \NH\ \about\ \E{23} \cs\ produce
near-optimal inversion and detected radiation in cloud-cloud amplification.
Proper motion measurements and comparisons of the disk and outflow masers offer
a most promising means to probe the structure and motion of TOR clouds.

To produce a TOR with the required angular width, clouds must maintain a column
density above \about\ \E{21} \cs\ during their rise to height $H \sim r$. If a
cloud expands at the speed of sound $c_{\rm s}$ while rising with velocity $v$,
then its size will increase by a factor $x \simeq 1 + (r/\Rc)(c_{\rm s}/v)$. We
expect $c_{\rm s}/v \simeq \E{-2}$ since $c_{\rm s}$ is 1\,\kms\ at a
temperature of 100 K and the velocity scale of the cloud outflow is expected to
be comparable to the Keplerian velocity, $v_{\rm K} =
208\,(\MBH/\rpc)^{1/2}$\,\kms.  With $\Rc/r \sim \E{-3}$, the 3-D expansion of
such a cloud during the rise to $H \sim r$ will cause its size to increase by a
factor $x$ \about\ 10 and its column density ($\propto \Mc/\Rc^2$) to decrease
by a factor $x^2$ \about\ 100, as required. Resistance to tidal shearing
dictates that the density inside the cloud increase as $1/r^3$ as it gets
closer to the central SBH (eq.\ \ref{eq:cloud}). A cloud of a given density can
exist only beyond a certain distance from the SBH; only denser clouds can
survive at smaller radii, leading to a radially stratified structure.
Similarly, the Keplerian velocity increases as $1/r^{1/2}$ as the SBH is
approached. Typical BLR cloud densities and velocities, discussed elsewhere in
these proceedings by H. Netzer, occur at $r$ \about\ \E{16}--\E{17} cm. It is
possible that the BLR inner boundary occurs where the clouds can no longer
overcome the SBH tidal shearing.

\subsection{TOR Outflow---Properties and Consequences}
\label{sec:outflow}

The kinetic luminosity of the TOR outflow is $L_{\rm k}^{\rm TOR} =
\int\frac12\Mc v_{\rm cl}^2\cdot\nc\vc\,dA$. The integration is over the disk
area where TOR clouds are injected out of the plane with vertical velocity
component \vc. This injection velocity can be estimated from the dispersion
velocity of molecular material in the disk --- an outflow can be considered to
have been established when the velocity of the ordered motion becomes
comparable to that of the local random motions. From the maser observations in
NGC~3079 \cite{Kondratko05} find that the velocity dispersion in a small region
($\le$ 5\x\E{16} cm) of strong emission is \about 14 \kms, which can be
considered typical of the local random motions. Therefore, we parameterize the
injection velocity as $\vc(r) = 10\,\kms\x v_6(\Rd) u(r)$, where $v_6(\Rd) =
\vc(\Rd)/10\,\kms$ is expected to be of order unity and where $u =
v(r)/v(\Rd)$. Without a full solution of the clumpy wind problem, which is far
from becoming available, the dimensionless profile $u(r)$ of velocity variation
across the disk plane remains unknown. However, it is quite certain that $u(r)$
decreases with $r$ from its maximum of $u(\Rd) = 1$; in particular, if the
turbulent velocity is a fraction of the local Keplerian velocity then $u =
(\Rd/r)^{1/2}$. With \Mc\nc = \mH\NH\Nc\ and \Rd\ from eq.\ \ref{eq:Rd} and
taking $T_{\rm sub}$ = 1500 K, the outflow kinetic luminosity is
\eq{
 L_{\rm k}^{\rm TOR}
      = 7\x\E{35} L_{45}^{1/2}N^{\rm tot}_{H,23} v_6^3(\Rd)\x\,I_3\ \erg.
}
Here $N^{\rm tot}_{H,23} = \No N_{H,23}$ is the column density through all
clouds in the outflow equatorial plane in units of \E{23}\,\cs\ and $I_3 =
\int_1^Y (u^3/y) dy\x Y/(Y - 1)$ is an unknown dimensionless factor of order
\la\ 1. This result shows that the kinetic luminosity of the TOR cloud outflow
is generally negligible in the overall energy budget of the AGN. This estimate
for $L_{\rm k}^{\rm TOR}$ does not lend support to the suggestion known as the
``AGN--Galaxy connection'' that outflows from the AGN may significantly affect
the host galaxy. However, although the energy outflow is not a significant
factor, the mass outflow is. A similar calculation for the mass outflow rate of
TOR clouds yields
\eq{\label{eq:Mdot1}
  \Mw = 0.02\,L_{45}^{1/2}\,N^{\rm tot}_{H,23}\,v_6(\Rd)\,\x\,I_1\ \Myr
}
where $I_1 = \int_1^Y (u/y) dy\x Y/(Y - 1)$ is another unknown factor of order
\la\ 1. Since the clouds are likely imbedded in a continuous wind, the actual
TOR outflow rate is probably somewhat higher. Steady-state mass conservation
requires accretion into the TOR with a rate $\MaTOR > \Mw$; if this constraint
is not met, the TOR mass is depleted within a few Keplerian orbits
\citep{Elitzur_Shlosman}. Mass accreted into the TOR and not injected into the
outflow finds its way into the black hole, and is the fuel for the AGN
luminosity. Therefore \MaTOR\ = \Mw\ + \MaBH, where \MaBH\ is the black hole
accretion rate, so $\MaBH = \alpha\MaTOR$ where $\alpha < 1$. In addition,
\MaBH\ is related to the AGN bolometric luminosity via $L = \eta\MaBH c^2$,
where $\eta \sim 0.1$ is the gravitational conversion efficiency. We end up
with $L = \epsilon\MaTOR c^2$, where $\epsilon = \eta\alpha$, and the accretion
into the TOR and the outflow away from it are related via
\eq{\label{eq:Mdot2}
   {\Mw\over\MaTOR} = {\epsilon\over L_{45}^{1/2}} N^{\rm tot}_{H,23}
                      v_6(\Rd)\x\,I_1.
}
This ratio cannot exceed unity, yet it increases when $L$ decreases with all
other factors remaining constant. Therefore, when the luminosity decreases the
bound $\Mw < \MaTOR$ is eventually violated, implying that the system cannot
sustain the cloud outflow rate inferred from the TOR properties. The wind
injection velocity $\vc(\Rd)$ is likely to change too when $L$ decreases
because the dust sublimation radius moves closer to the black hole (eq.\
\ref{eq:Rd}), where all velocities become larger. As with any location in the
accretion disk, the velocity scale of dynamic motions is determined by the
local Keplerian velocity $v_{\rm K}(\Rd) = (GM_\bullet/\Rd)^{1/2} \propto
(\LEdd/L^{1/2})^{1/2}$. If we assume that $\vc(\Rd)$ is proportional to $v_{\rm
K}(\Rd)$, eq.\ \ref{eq:Mdot2} becomes
\eq{\label{eq:Mdot3}
   {\Mw\over\MaTOR} = {\epsilon'\over L_{45}^{1/4}}
                      \left(\LEdd\over  L\right)^{1/2}
                  N^{\rm tot}_{H,23}
}
where the modified efficiency factor $\epsilon'$ absorbed all additional
proportionality constants, including $I_1$. At low accretion rates, the
accretion may switch to a radiatively inefficient mode \citep[RIAF; for recent
reviews see][]{Narayan02, Yuan07}, namely, the efficiency $\epsilon'$ decreases
too and we can expect the ratio $\epsilon'/L^{1/4}$ not to vary too much. The
TOR equatorial column density must remain $N^{\rm tot}_{H,23}$ \ga\ 1 as a
definition of the AGN torus, to provide the optical depth required by the
obscuration and observed IR emission. All in all, decreasing the luminosity at
a fixed black-hole mass (\LEdd) causes an increase of $\Mw/\MaTOR$, and when
this ratio exceeds unity the TOR disappears. {\em The torus should disappear in
low-luminosity AGN}. If we take $\epsilon \simeq 0.01$ and all other parameters
\about\ 1 in eq.\ \ref{eq:Mdot2}, the torus disappearance should occur at
bolometric luminosities below \about\ \E{42} \erg. Since the Eddington
luminosity of a \E7\Mo\ black-hole is \E{45} \erg, the torus disappearance can
be expected at a typical Eddington ratio $L/\LEdd$ \about\ \E{-3}.

\subsection{The AGN Low-Luminosity End}
\label{sec:lowL}

A key prediction of the wind scenario is that the torus disappears at low
bolometric luminosities ($\la$ \E{42} \erg) because mass accretion can no
longer sustain the required cloud outflow rate, i.e., the large column
densities. Observations seem to corroborate this prediction. In an HST study of
a complete sample of low-luminosity ($\la$ \E{42} \erg) FR I radio galaxies,
\cite{Chiaberge99} detected the compact core in 85\% of sources. Since the
radio selection is unbiased with respect to the AGN orientation, FR I sources
should contain similar numbers of type 1 and type 2 objects, and Chiaberge et
al suggested that the high detection rate of the central compact core implies
the absence of an obscuring torus. This suggestion was corroborated by
\cite{Whys04} who demonstrated the absence of a dusty torus in M87, one of the
sources in the FR I sample, by placing stringent limits on its thermal IR
emission. Observations by \cite{Perlman07} further solidified this conclusion.
The IR radiation they detect in the M87 core comes primarily from non-thermal
jet emission, with only a trace of thermal emission that is much weaker than
what would be expected from an AGN torus and that can be attributed to
neighboring dust. The behavior displayed by M87 appears to be common in FR I
sources. Van der Wolk et al report in these proceedings the results of high
resolution 12 \mic\ imaging observations of the nuclei of 27 radio galaxies,
performed with the VISIR instrument on the VLT. The reported observations
provide strong confirmation of the torus disappearance in FR I sources.  They
show that all the FR I objects in the sample lack dusty torus thermal emission,
although they have non-thermal nuclei.  Thermal dust emission was detected in
about half the FR II nuclei, which generally have higher luminosities. In
contrast, in almost all broad line radio galaxies in the sample the
observations detected the thermal nucleus. Significantly, \cite{Ogle07} find
that most FR I and half of FR II sources have $L/\LEdd < 3\cdot\E{-3}$, while
all sources with broad Balmer lines have $L/\LEdd > 3\cdot\E{-3}$.

The low-luminosities sources known as LINERs provide additional evidence for
the torus disappearance. \cite{Maoz05} conducted UV monitoring of LINERs with
$L$ \la\ \E{42} \erg\ and detected variability in most of them.  This
demonstrates that the AGN makes a significant contribution to the UV radiation
in each of the monitored sources and that it is relatively unobscured in all
the observed LINERs, which included both type 1 and type 2. Furthermore, the
histograms of UV colors of the type 1 and 2 LINERs show an overlap between the
two populations, with the difference between the histogram peaks corresponding
to dust obscuration in the type 2 objects of only \about\ 1 magnitude in the R
band. Such toroidal obscuration is minute in comparison with the torus
obscuration in higher luminosity AGN. The predicted torus disappearance at low
$L$ does not imply that the cloud component of the disk wind is abruptly
extinguished, only that its outflow rate is less than required by the IR
emission observed in quasars and high-luminosity Seyferts. When \Mw\ drops
below these ``standard'' torus values, the outflow still provides toroidal
obscuration as long as its column exceeds \about\ \E{21} \cs. Indeed, Maoz et
al find that some LINERs do have obscuration, but much smaller than
``standard''. Line transmission through a low-obscuration torus might also
explain the low polarizations of broad H$\alpha$ lines observed by
\cite{Barth99} in some low luminosity systems.

If the toroidal obscuration were the only component removed from the system,
all low luminosity AGN would become type 1 sources. In fact, among the LINERs
monitored and found to be variable by Maoz et al there were both sources with
broad H$\alpha$ wings (type 1) and those without (type 2). Since all objects
are relatively unobscured, the broad line component is truly missing in the
type 2 sources in this sample. Similarly, \cite{Panessa02} note that the BLR is
weak or absent in low luminosity AGN, and \cite{Laor03} presents arguments that
some ``true'' type 2 sources, i.e., having no obscured BLR, do exist among AGNs
with $L$ \la\ \E{42}\,\erg. The absence of broad lines in these sources cannot
be attributed to toroidal obscuration because their X-ray emission is largely
unobscured. These findings have a simple explanation if when $L$ decreases
further beyond the disappearance of the TOR outflow, the suppression of mass
outflow spreads radially inward from the disk's dusty, molecular region into
its atomic, ionized zone. Then the torus disappearance, i.e., removal of the
toroidal obscuration by the dusty wind, would be followed by a diminished
outflow from the inner ionized zone and disappearance of the BLR at some lower
luminosity. Indeed, \cite{Ogle07} find that all sources with broad Balmer lines
have $L/\LEdd\ \ga\ {3\cdot\E{-3}}$, as noted above. Such an inward progression
of the outflow turnoff as the accretion rate decreases can be expected
naturally in the context of disk winds because mass outflow increases with the
disk area. A diminished supply of accreted mass may suffice to support an
outflow from the inner parts of the disk but not from the larger area of its
outer regions. And with further decrease in inflow rate, even the smaller inner
area cannot sustain the disk outflow. Since the accreted mass cannot be
channeled in full into the central black hole, the system must find another
channel for release of the excess accreted mass, and the only one remaining is
the radio jets. Indeed, \cite{Ho02} finds that the AGN radio loudness $\R =
L_{\rm radio}/L_{\rm opt}$ is {\em inversely} correlated with the mass
accretion rate $L/\LEdd$, a correlation further verified by \cite*{Greene06}.
This finding is supported by \cite{Sikora07}, who have greatly expanded this
correlation and found an intriguing result: \R\ indeed increases inversely with
$L/\LEdd$, but only so long as $L/\LEdd$ remains \ga\ \E{-3}. At smaller
accretion rates, which include all FR I radio galaxies, the radio loudness
saturates and remains constant at \R\ \about \E4. This is precisely the
expected behavior if as the outflow diminishes, the jets are fed an
increasingly larger fraction of the accreted mass and finally, once the outflow
is extinguished, all the inflowing material not funneled into the black hole is
channeled into the jets, whose feeding thus saturates at a high conversion
efficiency of accreted mass. It is important to note that radio loudness
reflects the relative contribution of radio to the overall radiative emission;
a source can be radio loud even at a low level of radio emission if its overall
luminosity is small, and vice versa.

\begin{figure}[ht]
 \centering\leavevmode\includegraphics[width=0.6\hsize,clip]{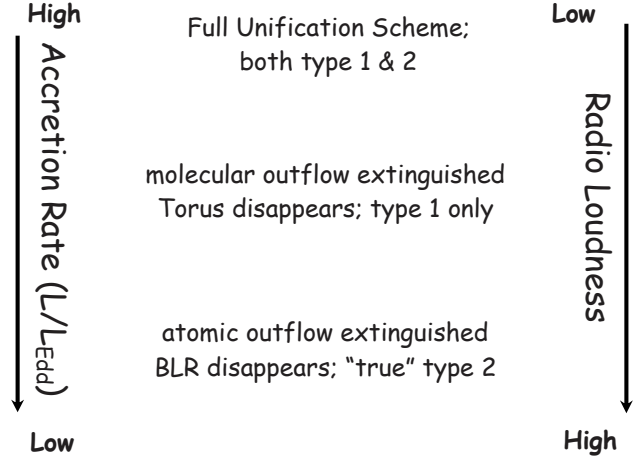}
\caption{Conjectured scheme for AGN evolution with decreasing accretion rate.}
\label{Fig:AGN_scheme}
\end{figure}

The evolutionary scheme just outlines is sketched in figure
\ref{Fig:AGN_scheme}. A similar anti-correlation between radio loudness and
accretion rate exists also in X-ray binaries. These sources display switches
between radio quiet states of high X-ray emission and radio loud states with
low X-ray emission \citep{Fender04}. While in X-ray binaries this behavior can
be followed with time in a given source, in AGN it is only manifested
statistically. Comparative studies of AGN and X-ray binaries seem to be a most
useful avenue to pursue.

\subsection{Limitations}

The scheme just outlined was not derived from a detailed dynamical model for
clumpy disk outflow in AGN. Such models do not yet exist. Instead, the results
follow from considerations of basic conservation laws within the framework of
the wind paradigm for the AGN torus. The conclusion that the TOR should
disappear at low accretion rates  is almost unavoidable because the outflow
rate decreases more slowly than does the luminosity. In eq.\ \ref{eq:Mdot2},
our ignorance of all dynamics details is lumped into the efficiency factor
$\epsilon$. Although this factor is largely unknown, the torus disappearance
conclusion would be off only if $\epsilon$ varied with luminosity in a way that
completely offset the explicit $L$-dependencies in eqs.\ \ref{eq:Mdot2} and
\ref{eq:Mdot3}. Observations indicate that this is not the case. The most
recent review by \cite{Ho08} presents extensive observational evidence for the
disappearance of the torus and the BLR in low luminosity AGN.

The lack of explicit dynamics model is both the weakness and strength of the
presented arguments. Our results provide constraints that have to be met by any
detailed model constructed to explain the AGN torus within the disk wind
paradigm. It should be noted that we were able to estimate only the properties
of the TOR outflow and deduce that this outflow must be turned off at low
luminosities. The BLR disappearance is then deduced from the observational
evidence that the low luminosity AGN do not become exclusively type 1 objects.
A possible dynamical explanation is offered by \cite{Nicastro00}, whose work on
the role of radiation pressure provides the earliest prediction for BLR
disappearance at low $L/\LEdd$.

\subsection{Final Speculations}

While outflow considerations set limits on the AGN low luminosity end, they do
not provide any constraints on the TOR at high luminosities. From observations,
the fraction $f_2$ of obscured AGN decreases with bolometric luminosity (see
sec.\ \ref{sec:f2}).  As is evident from figure \ref{Fig:f2}, this can be
produced by either a decrease of $\sigma$ at constant \No\ or a decrease of
\No\ at constant $\sigma$. The torus radial dimensions are set from its inner
radius \Rd, which increases with $L$ due to dust sublimation (eq.\
\ref{eq:Rd}). Therefore, the decreasing-$\sigma$ option would arise if the TOR
height were independent of luminosity or increased more slowly than $L^{1/2}$,
the decreasing-\No\ option would arise if the TOR outer radius were independent
of luminosity or increased more slowly than $L^{1/2}$. The outflow outer radius
is plausibly confined to within the SBH gravitational influence radius, \Ro\
$<$ 35\,pc\x$(\MBH/\Omega_1^2)^{1/3}$ (sec.\ 5). As noted in sec.\
\ref{sec:size}, an upper bound on the torus outer radius set conservatively
from IR observations is $\Ro < 12\,L_{45}^{1/2}$ pc. So whenever $L_{45}\ \ga\
9(\MBH/\Omega_1^2)^{2/3}$, the TOR outflow is cut at an outer radius smaller
than that containing the typical number of clouds according to IR observations,
providing a possible explanation for the decreasing-\No\ option.

Finally, the AGN bolometric luminosity implies accretion into the black hole at
the rate $\MaBH \sim 0.1 L_{45}\, \Myr$, assuming radiation efficiency $\eta$
\about 0.1 (see sec.\ \ref{sec:outflow}). The AGN phase lasts probably \about
\E7--\E8\ years, thus the overall mass accreted is \about\ \E6--\E7$L_{45}$
\Mo. Since Seyfert luminosities do not exceed \about \E{45} \erg\ at most, the
entire mass they typically accrete from the host galaxy during an AGN episode
is no more than that contained in a single giant molecular cloud. In contrast,
QSO luminosities reach upward of \E{48} \erg. If QSO lifetimes are similar, the
mass fed into the SBH is comparable to that of an entire galaxy. This suggests
that the quasar and Seyfert nuclear activities could be triggered differently.
Starting a quasar might require interaction with another galaxy to trigger
accretion of the prodigious amounts of fuel that can only be supplied by a full
galactic mass. In contrast, a Seyfert episode may represent the occasional
capture of a giant molecular cloud by the SBH that is now known to exist at the
center of most galaxies.

\section{Concluding Remarks}

The clumpy nature of the obscuring torus, long ago deduced from theoretical
considerations and now being verified by observations, dictates a fundamental
change in the approach to some basic issues regarding AGN. Foremost among them
is analysis of the type 1 and type 2 source statistics. Because the viewing
angle determines the obscuration only in the sense of probability, and not
absolutely, translating source statistics into torus parameters becomes a
difficult task. In addition to the torus angular width $\sigma$, which has been
used as the sole parameter of such analysis, it involves also the mean number
of clouds \No, making it impossible to determine both of them from the single
measured quantity $f_2$. In particular, the observed decrease of $f_2$ with
increasing luminosity can be attributed to a decrease of either $\sigma$ or
\No\ or both. Meaningful constraints on the torus physical properties can be
placed only through combination of different types of data. Comparisons of
UV/optical and X-ray properties in individual sources and in samples offer
especially promising analysis modes for deducing detailed torus properties.

The low-luminosity end of AGN offers a rich variety of interesting phenomena.
The consecutive disappearances of the torus and the BLR at low luminosities, or
low Eddington ratios, and the simultaneous increase in radio loudness fit
naturally into the outflow paradigm for the AGN torus. Detailed observational
and theoretical studies of these phenomena would enhance significantly our
understanding of the black-hole environment.

\textbf{Acknowledgements:} Discussions with numerous colleagues were
instrumental in formulating many of the ideas presented here. Special thanks to
Roberto Maiolino, who pointed out the important role of the Eddington ratio
described in sec. \ref{sec:outflow}, Ari Laor and Isaac Shlosman. Partial
support by NSF and NASA is gratefully acknowledged.

\def\aj{AJ}%
\def\actaa{Acta Astron.}%
\def\araa{ARA\&A}%
\def\apj{ApJ}%
\def\apjl{ApJ}%
\def\apjs{ApJS}%
\def\ao{Appl.~Opt.}%
\def\apss{Ap\&SS}%
\def\aap{A\&A}%
\def\aapr{A\&A~Rev.}%
\def\aaps{A\&AS}%
\def\azh{AZh}%
\def\baas{BAAS}%
\def\bac{Bull. astr. Inst. Czechosl.}%
\def\caa{Chinese Astron. Astrophys.}%
\def\cjaa{Chinese J. Astron. Astrophys.}%
\def\icarus{Icarus}%
\def\jcap{J. Cosmology Astropart. Phys.}%
\def\jrasc{JRASC}%
\def\mnras{MNRAS}%
\def\memras{MmRAS}%
\def\na{New A}%
\def\nar{New A Rev.}%
\def\pasa{PASA}%
\def\pra{Phys.~Rev.~A}%
\def\prb{Phys.~Rev.~B}%
\def\prc{Phys.~Rev.~C}%
\def\prd{Phys.~Rev.~D}%
\def\pre{Phys.~Rev.~E}%
\def\prl{Phys.~Rev.~Lett.}%
\def\pasp{PASP}%
\def\pasj{PASJ}%
\def\qjras{QJRAS}%
\def\rmxaa{Rev. Mexicana Astron. Astrofis.}%
\def\skytel{S\&T}%
\def\solphys{Sol.~Phys.}%
\def\sovast{Soviet~Ast.}%
\def\ssr{Space~Sci.~Rev.}%
\def\zap{ZAp}%
\def\nat{Nature}%
\def\iaucirc{IAU~Circ.}%
\def\aplett{Astrophys.~Lett.}%
\def\apspr{Astrophys.~Space~Phys.~Res.}%
\def\bain{Bull.~Astron.~Inst.~Netherlands}%
\def\fcp{Fund.~Cosmic~Phys.}%
\def\gca{Geochim.~Cosmochim.~Acta}%
\def\grl{Geophys.~Res.~Lett.}%
\def\jcp{J.~Chem.~Phys.}%
\def\jgr{J.~Geophys.~Res.}%
\def\jqsrt{J.~Quant.~Spec.~Radiat.~Transf.}%
\def\memsai{Mem.~Soc.~Astron.~Italiana}%
\def\nphysa{Nucl.~Phys.~A}%
\def\physrep{Phys.~Rep.}%
\def\physscr{Phys.~Scr}%
\def\planss{Planet.~Space~Sci.}%
\def\procspie{Proc.~SPIE}%
\let\astap=\aap
\let\apjlett=\apjl
\let\apjsupp=\apjs
\let\applopt=\ao
%


\begin{thebibliography}{109}
\expandafter\ifx\csname natexlab\endcsname\relax\def\natexlab#1{#1}\fi

\bibitem[{{Alonso-Herrero} {et~al.}(2003){Alonso-Herrero}, {Quillen}, {Rieke},
  {Ivanov}, \& {Efstathiou}}]{Almudena03}
{Alonso-Herrero}, A., {Quillen}, A.~C., {Rieke}, G.~H., {Ivanov}, V.~D., \&
  {Efstathiou}, A. 2003, \aj, 126, 81

\bibitem[{{Aretxaga} {et~al.}(1999){Aretxaga}, {Joguet}, {Kunth}, {Melnick}, \&
  {Terlevich}}]{Aret99}
{Aretxaga}, I., {Joguet}, B., {Kunth}, D., {Melnick}, J., \& {Terlevich}, R.~J.
  1999, \apjl, 519, L123

\bibitem[{{Barth} {et~al.}(1999){Barth}, {Filippenko}, \& {Moran}}]{Barth99}
{Barth}, A.~J., {Filippenko}, A.~V., \& {Moran}, E.~C. 1999, \apj, 525, 673

\bibitem[{{Bassani} {et~al.}(1999){Bassani}, {Dadina}, {Maiolino}, {Salvati},
  {Risaliti}, {della Ceca}, {Matt}, \& {Zamorani}}]{Bassani99}
{Bassani}, L., {Dadina}, M., {Maiolino}, R., {et~al.} 1999, \apjs, 121, 473

\bibitem[{{Blandford} \& {Payne}(1982)}]{Blandford_Payne}
{Blandford}, R.~D. \& {Payne}, D.~G. 1982, \mnras, 199, 883

\bibitem[{{Bottorff} {et~al.}(1997){Bottorff}, {Korista}, {Shlosman}, \&
  {Blandford}}]{Bottorff97}
{Bottorff}, M., {Korista}, K.~T., {Shlosman}, I., \& {Blandford}, R.~D. 1997,
  \apj, 479, 200

\bibitem[{{Bottorff} {et~al.}(2000){Bottorff}, {Korista}, \&
  {Shlosman}}]{Bottorff00}
{Bottorff}, M.~C., {Korista}, K.~T., \& {Shlosman}, I. 2000, \apj, 537, 134

\bibitem[{{Braito} {et~al.}(2004){Braito}, {Della Ceca}, {Piconcelli},
  {Severgnini}, {Bassani}, {Cappi}, {Franceschini}, {Iwasawa}, {Malaguti},
  {Marziani}, {Palumbo}, {Persic}, {Risaliti}, \& {Salvati}}]{Braito04}
{Braito}, V., {Della Ceca}, R., {Piconcelli}, E., {et~al.} 2004, \aap, 420, 79

\bibitem[{{Buchanan} {et~al.}(2006){Buchanan}, {Gallimore}, {O'Dea}, {Baum},
  {Axon}, {Robinson}, {Elitzur}, \& {Elvis}}]{Buchanan06}
{Buchanan}, C.~L., {Gallimore}, J.~F., {O'Dea}, C.~P., {et~al.} 2006, \aj, 132,
  401

\bibitem[{{Chiaberge} {et~al.}(1999){Chiaberge}, {Capetti}, \&
  {Celotti}}]{Chiaberge99}
{Chiaberge}, M., {Capetti}, A., \& {Celotti}, A. 1999, \aap, 349, 77

\bibitem[{{Conway} {et~al.}(2005){Conway}, {Elitzur}, \& {Parra}}]{Conway05}
{Conway}, J., {Elitzur}, M., \& {Parra}, R. 2005, \apss, 295, 319

\bibitem[{{Conway} {et~al.}(2008){Conway}, {Elitzur}, \& {Parra}}]{Conway08}
{Conway}, J., {Elitzur}, M., \& {Parra}, R. 2008, in preparation

\bibitem[{{Crenshaw} \& {Kraemer}(2000)}]{Crenshaw00}
{Crenshaw}, D.~M. \& {Kraemer}, S.~B. 2000, \apjl, 532, L101

\bibitem[{{Draine}(2003)}]{Draine03}
{Draine}, B.~T. 2003, \apj, 598, 1017

\bibitem[{{Eckart} {et~al.}(2006){Eckart}, {Stern}, {Helfand}, {Harrison},
  {Mao}, \& {Yost}}]{Eckart06}
{Eckart}, M.~E., {Stern}, D., {Helfand}, D.~J., {et~al.} 2006, \apjs, 165, 19

\bibitem[{{Efstathiou} \& {Rowan-Robinson}(1994)}]{EfRR94}
{Efstathiou}, A. \& {Rowan-Robinson}, M. 1994, \mnras, 266, 212

\bibitem[{{Elitzur} {et~al.}(1989){Elitzur}, {Hollenbach}, \& {McKee}}]{EHM89}
{Elitzur}, M., {Hollenbach}, D.~J., \& {McKee}, C.~F. 1989, \apj, 346, 983

\bibitem[{{Elitzur} \& {Shlosman}(2006)}]{Elitzur_Shlosman}
{Elitzur}, M. \& {Shlosman}, I. 2006, \apjl, 648, L101

\bibitem[{{Elvis}(2004)}]{Elvis_winds}
{Elvis}, M. 2004, in ASP Conf. Ser. 311: AGN Physics with the Sloan Digital Sky
  Survey, ed. G.~T. {Richards} \& P.~B. {Hall}, 109

\bibitem[{{Elvis} {et~al.}(2004){Elvis}, {Risaliti}, {Nicastro}, {Miller},
  {Fiore}, \& {Puccetti}}]{Elvis04}
{Elvis}, M., {Risaliti}, G., {Nicastro}, F., {et~al.} 2004, \apjl, 615, L25

\bibitem[{{Emmering} {et~al.}(1992){Emmering}, {Blandford}, \&
  {Shlosman}}]{Emmering92}
{Emmering}, R.~T., {Blandford}, R.~D., \& {Shlosman}, I. 1992, \apj, 385, 460

\bibitem[{{Eracleous}(2004)}]{Eracleous04}
{Eracleous}, M. 2004, in ASP Conf. Ser. 311: AGN Physics with the Sloan Digital
  Sky Survey, ed. G.~T. {Richards} \& P.~B. {Hall}, 183

\bibitem[{{Everett}(2005)}]{Everett05}
{Everett}, J.~E. 2005, \apj, 631, 689

\bibitem[{{Fender} {et~al.}(2004){Fender}, {Belloni}, \& {Gallo}}]{Fender04}
{Fender}, R.~P., {Belloni}, T.~M., \& {Gallo}, E. 2004, \mnras, 355, 1105

\bibitem[{{Fritz} {et~al.}(2006){Fritz}, {Franceschini}, \&
  {Hatziminaoglou}}]{Fritz06}
{Fritz}, J., {Franceschini}, A., \& {Hatziminaoglou}, E. 2006, \mnras, 366, 767

\bibitem[{{Gallagher} {et~al.}(2006){Gallagher}, {Brandt}, {Chartas},
  {Priddey}, {Garmire}, \& {Sambruna}}]{Gallagher06}
{Gallagher}, S.~C., {Brandt}, W.~N., {Chartas}, G., {et~al.} 2006, \apj, 644,
  709

\bibitem[{{Gallagher} \& {Everett}(2007)}]{Gallagher07}
{Gallagher}, S.~C. \& {Everett}, J.~E. 2007, in ASP Conf. Ser. 373: The Central
  Engine of Active Galactic Nuclei, ed. L.~C. {Ho} \& J.-M. {Wang}, 305--314

\bibitem[{{Galliano} {et~al.}(2003){Galliano}, {Alloin}, {Granato}, \&
  {Villar-Mart{\'{\i}}n}}]{Galliano03}
{Galliano}, E., {Alloin}, D., {Granato}, G.~L., \& {Villar-Mart{\'{\i}}n}, M.
  2003, \aap, 412, 615

\bibitem[{{Gallimore} {et~al.}(2001){Gallimore}, {Henkel}, {Baum}, {Glass},
  {Claussen}, {Prieto}, \& {Von Kap-herr}}]{Gallimore01}
{Gallimore}, J.~F., {Henkel}, C., {Baum}, S.~A., {et~al.} 2001, \apj, 556, 694

\bibitem[{{Garcet} {et~al.}(2007){Garcet}, {Gandhi}, {Gosset}, {Sprimont},
  {Surdej}, {Borkowski}, {Tajer}, {Pacaud}, {Pierre}, {Chiappetti}, {Maccagni},
  {Page}, {Carrera}, {Tedds}, {Mateos}, {Krumpe}, {Contini}, {Corral},
  {Ebrero}, {Gavignaud}, {Schwope}, {Le F{\`e}vre}, {Polletta}, {Rosen},
  {Lonsdale}, {Watson}, {Borczyk}, \& {Vaisanen}}]{Garcet07}
{Garcet}, O., {Gandhi}, P., {Gosset}, E., {et~al.} 2007, \aap, 474, 473

\bibitem[{{George} {et~al.}(1998){George}, {Turner}, {Netzer}, {Nandra},
  {Mushotzky}, \& {Yaqoob}}]{George98}
{George}, I.~M., {Turner}, T.~J., {Netzer}, H., {et~al.} 1998, \apjs, 114, 73

\bibitem[{{Granato} \& {Danese}(1994)}]{Granato94}
{Granato}, G.~L. \& {Danese}, L. 1994, \mnras, 268, 235

\bibitem[{{Granato} {et~al.}(1997){Granato}, {Danese}, \&
  {Franceschini}}]{Granato97}
{Granato}, G.~L., {Danese}, L., \& {Franceschini}, A. 1997, \apj, 486, 147

\bibitem[{{Gratadour} {et~al.}(2003){Gratadour}, {Cl{\'e}net}, {Rouan}, {Lai},
  \& {Forveille}}]{Gratadour03}
{Gratadour}, D., {Cl{\'e}net}, Y., {Rouan}, D., {Lai}, O., \& {Forveille}, T.
  2003, \aap, 411, 335

\bibitem[{{Greene} {et~al.}(2006){Greene}, {Ho}, \& {Ulvestad}}]{Greene06}
{Greene}, J.~E., {Ho}, L.~C., \& {Ulvestad}, J.~S. 2006, \apj, 636, 56

\bibitem[{{Greenhill}(2005)}]{Greenhill05}
{Greenhill}, L.~J. 2005, in ASP Conf. Ser. 340: Future Directions in High
  Resolution Astronomy, ed. J.~{Romney} \& M.~{Reid}, 203

\bibitem[{{Greenhill} {et~al.}(2003){Greenhill}, {Booth}, {Ellingsen},
  {Herrnstein}, {Jauncey}, {McCulloch}, {Moran}, {Norris}, {Reynolds}, \&
  {Tzioumis}}]{Greenhill03}
{Greenhill}, L.~J., {Booth}, R.~S., {Ellingsen}, S.~P., {et~al.} 2003, \apj,
  590, 162

\bibitem[{{Greenhill} \& {Gwinn}(1997)}]{Greenhill97}
{Greenhill}, L.~J. \& {Gwinn}, C.~R. 1997, \apss, 248, 261

\bibitem[{{Guainazzi} {et~al.}(2001){Guainazzi}, {Fiore}, {Matt}, \&
  {Perola}}]{Guainazzi01}
{Guainazzi}, M., {Fiore}, F., {Matt}, G., \& {Perola}, G.~C. 2001, \mnras, 327,
  323

\bibitem[{{Guainazzi} {et~al.}(2005){Guainazzi}, {Matt}, \&
  {Perola}}]{Guainazzi05}
{Guainazzi}, M., {Matt}, G., \& {Perola}, G.~C. 2005, \aap, 444, 119

\bibitem[{{Hall} {et~al.}(2003){Hall}, {Hutsem{\'e}kers}, {Anderson},
  {Brinkmann}, {Fan}, {Schneider}, \& {York}}]{Hall03}
{Hall}, P.~B., {Hutsem{\'e}kers}, D., {Anderson}, S.~F., {et~al.} 2003, \apj,
  593, 189

\bibitem[{{Hao} {et~al.}(2005){Hao}, {Strauss}, {Fan}, {Tremonti}, {Schlegel},
  {Heckman}, {Kauffmann}, {Blanton}, {Gunn}, {Hall}, {Ivezi{\'c}}, {Knapp},
  {Krolik}, {Lupton}, {Richards}, {Schneider}, {Strateva}, {Zakamska},
  {Brinkmann}, \& {Szokoly}}]{Hao05b}
{Hao}, L., {Strauss}, M.~A., {Fan}, X., {et~al.} 2005, \aj, 129, 1795

\bibitem[{{Hao} {et~al.}(2007){Hao}, {Weedman}, {Spoon}, {Marshall},
  {Levenson}, {Elitzur}, \& {Houck}}]{Hao07}
{Hao}, L., {Weedman}, D.~W., {Spoon}, H.~W.~W., {et~al.} 2007, \apjl, 655, L77

\bibitem[{{Ho}(2002)}]{Ho02}
{Ho}, L.~C. 2002, \apj, 564, 120

\bibitem[{{Ho}(2008)}]{Ho08}
{Ho}, L.~C. 2008, \araa, in press (arXiv0803.2268)

\bibitem[{{Horst} {et~al.}(2006){Horst}, {Smette}, {Gandhi}, \&
  {Duschl}}]{Horst06}
{Horst}, H., {Smette}, A., {Gandhi}, P., \& {Duschl}, W.~J. 2006, \aap, 457,
  L17

\bibitem[{{Ivezi{\'c}} \& {Elitzur}(1997)}]{IE97}
{Ivezi{\'c}}, {\v Z}. \& {Elitzur}, M. 1997, \mnras, 287, 799

\bibitem[{{Jaffe} {et~al.}(2004){Jaffe}, {Meisenheimer}, {R{\"o}ttgering},
  {Leinert}, {Richichi}, {Chesneau}, {Fraix-Burnet}, {Glazenborg-Kluttig},
  {Granato}, {Graser}, {Heijligers}, {K{\"o}hler}, {Malbet}, {Miley},
  {Paresce}, {Pel}, {Perrin}, {Przygodda}, {Schoeller}, {Sol}, {Waters},
  {Weigelt}, {Woillez}, \& {de Zeeuw}}]{Jaffe04}
{Jaffe}, W., {Meisenheimer}, K., {R{\"o}ttgering}, H.~J.~A., {et~al.} 2004,
  \nat, 429, 47

\bibitem[{{Kartje} \& {K{\"o}nigl}(1996)}]{Kartje96}
{Kartje}, J.~F. \& {K{\"o}nigl}, A. 1996, Vistas in Astronomy, 40, 133

\bibitem[{{Kartje} {et~al.}(1999){Kartje}, {K{\"o}nigl}, \&
  {Elitzur}}]{Kartje99}
{Kartje}, J.~F., {K{\"o}nigl}, A., \& {Elitzur}, M. 1999, \apj, 513, 180

\bibitem[{{Kinney} {et~al.}(2000){Kinney}, {Schmitt}, {Clarke}, {Pringle},
  {Ulvestad}, \& {Antonucci}}]{Kinney00}
{Kinney}, A.~L., {Schmitt}, H.~R., {Clarke}, C.~J., {et~al.} 2000, \apj, 537,
  152

\bibitem[{{Kondratko} {et~al.}(2005){Kondratko}, {Greenhill}, \&
  {Moran}}]{Kondratko05}
{Kondratko}, P.~T., {Greenhill}, L.~J., \& {Moran}, J.~M. 2005, \apj, 618, 618

\bibitem[{{K{\"o}nigl} \& {Kartje}(1994)}]{Konigl94}
{K{\"o}nigl}, A. \& {Kartje}, J.~F. 1994, \apj, 434, 446

\bibitem[{{Krolik} \& {Begelman}(1988)}]{Krolik88}
{Krolik}, J.~H. \& {Begelman}, M.~C. 1988, \apj, 329, 702

\bibitem[{{Krolik} {et~al.}(1994){Krolik}, {Madau}, \& {Zycki}}]{Krolik94}
{Krolik}, J.~H., {Madau}, P., \& {Zycki}, P.~T. 1994, \apjl, 420, L57

\bibitem[{{Laor}(2003)}]{Laor03}
{Laor}, A. 2003, \apj, 590, 86

\bibitem[{{Levenson} {et~al.}(2002){Levenson}, {Krolik}, {{\.Z}ycki},
  {Heckman}, {Weaver}, {Awaki}, \& {Terashima}}]{Levenson02}
{Levenson}, N.~A., {Krolik}, J.~H., {{\.Z}ycki}, P.~T., {et~al.} 2002, \apjl,
  573, L81

\bibitem[{{Levenson} {et~al.}(2007){Levenson}, {Sirocky}, {Hao}, {Spoon},
  {Marshall}, {Elitzur}, \& {Houck}}]{Levenson07}
{Levenson}, N.~A., {Sirocky}, M.~M., {Hao}, L., {et~al.} 2007, \apjl, 654, L45

\bibitem[{{Lutz} {et~al.}(2004){Lutz}, {Maiolino}, {Spoon}, \&
  {Moorwood}}]{Lutz04}
{Lutz}, D., {Maiolino}, R., {Spoon}, H.~W.~W., \& {Moorwood}, A.~F.~M. 2004,
  \aap, 418, 465

\bibitem[{{Maccacaro} {et~al.}(1982){Maccacaro}, {Perola}, \&
  {Elvis}}]{Maccacaro82}
{Maccacaro}, T., {Perola}, G.~C., \& {Elvis}, M. 1982, \apj, 257, 47

\bibitem[{{Maiolino} {et~al.}(2001){Maiolino}, {Marconi}, {Salvati},
  {Risaliti}, {Severgnini}, {Oliva}, {La Franca}, \& {Vanzi}}]{Maiolino01}
{Maiolino}, R., {Marconi}, A., {Salvati}, M., {et~al.} 2001, \aap, 365, 28

\bibitem[{{Maiolino} {et~al.}(2007){Maiolino}, {Shemmer}, {Imanishi}, {Netzer},
  {Oliva}, {Lutz}, \& {Sturm}}]{Maiolino07a}
{Maiolino}, R., {Shemmer}, O., {Imanishi}, M., {et~al.} 2007, \aap, 468, 979

\bibitem[{{Maoz} {et~al.}(2005){Maoz}, {Nagar}, {Falcke}, \& {Wilson}}]{Maoz05}
{Maoz}, D., {Nagar}, N.~M., {Falcke}, H., \& {Wilson}, A.~S. 2005, \apj, 625,
  699

\bibitem[{{Mason} {et~al.}(2006){Mason}, {Geballe}, {Packham}, {Levenson},
  {Elitzur}, {Fisher}, \& {Perlman}}]{Mason06}
{Mason}, R.~E., {Geballe}, T.~R., {Packham}, C., {et~al.} 2006, \apj, 640, 612

\bibitem[{{Mathis} {et~al.}(1977){Mathis}, {Rumpl}, \& {Nordsieck}}]{MRN}
{Mathis}, J.~S., {Rumpl}, W., \& {Nordsieck}, K.~H. 1977, \apj, 217, 425

\bibitem[{{Nandra} \& {Pounds}(1994)}]{Nandra94}
{Nandra}, K. \& {Pounds}, K.~A. 1994, \mnras, 268, 405

\bibitem[{{Narayan}(2002)}]{Narayan02}
{Narayan}, R. 2002, in Lighthouses of the Universe: The Most Luminous Celestial
  Objects and Their Use for Cosmology, ed. M.~{Gilfanov}, R.~{Sunyeav}, \&
  E.~{Churazov}, 405

\bibitem[{{Natta} \& {Panagia}(1984)}]{Natta84}
{Natta}, A. \& {Panagia}, N. 1984, \apj, 287, 228

\bibitem[{{Nenkova} {et~al.}(2002){Nenkova}, {Ivezi{\'c}}, \&
  {Elitzur}}]{NIE02}
{Nenkova}, M., {Ivezi{\'c}}, {\v Z}., \& {Elitzur}, M. 2002, \apjl, 570, L9

\bibitem[{{Nenkova} {et~al.}(2008{\natexlab{a}}){Nenkova}, Sirocky,
  {Ivezi{\'c}}, \& {Elitzur}}]{AGN1}
{Nenkova}, M., Sirocky, M.~M., {Ivezi{\'c}}, {\v Z}., \& {Elitzur}, M.
  2008{\natexlab{a}}, \apj, submitted

\bibitem[{{Nenkova} {et~al.}(2008{\natexlab{b}}){Nenkova}, Sirocky, Nikutta,
  {Ivezi{\'c}}, \& {Elitzur}}]{AGN2}
{Nenkova}, M., Sirocky, M.~M., Nikutta, R., {Ivezi{\'c}}, {\v Z}., \&
  {Elitzur}, M. 2008{\natexlab{b}}, \apj, submitted

\bibitem[{{Netzer} \& {Laor}(1993)}]{Netzer_Laor}
{Netzer}, H. \& {Laor}, A. 1993, \apjl, 404, L51

\bibitem[{{Nicastro}(2000)}]{Nicastro00}
{Nicastro}, F. 2000, \apjl, 530, L65

\bibitem[{{Ogle} {et~al.}(2007){Ogle}, {Antonucci}, \& {Whysong}}]{Ogle07}
{Ogle}, P.~M., {Antonucci}, R.~R.~J., \& {Whysong}, D. 2007, in ASP Conf. Ser.
  373: The Central Engine of Active Galactic Nuclei, ed. L.~C. {Ho} \& J.-M.
  {Wang}, 578--581

\bibitem[{{Ossenkopf} {et~al.}(1992){Ossenkopf}, {Henning}, \&
  {Mathis}}]{Ossenkopf92}
{Ossenkopf}, V., {Henning}, T., \& {Mathis}, J.~S. 1992, \aap, 261, 567

\bibitem[{{Packham} {et~al.}(2007){Packham}, {Young}, {Fisher}, {Volk},
  {Mason}, {Hough}, {Roche}, {Elitzur}, {Radomski}, \& {Perlman}}]{Packham07}
{Packham}, C., {Young}, S., {Fisher}, S., {et~al.} 2007, \apjl, 661, L29

\bibitem[{{Page} {et~al.}(2006){Page}, {Loaring}, {Dwelly}, {Mason}, {McHardy},
  {Gunn}, {Moss}, {Sasseen}, {Cordova}, {Kennea}, \& {Seymour}}]{Page06}
{Page}, M.~J., {Loaring}, N.~S., {Dwelly}, T., {et~al.} 2006, \mnras, 369, 156

\bibitem[{{Panessa} \& {Bassani}(2002)}]{Panessa02}
{Panessa}, F. \& {Bassani}, L. 2002, \aap, 394, 435

\bibitem[{{Perlman} {et~al.}(2007){Perlman}, {Mason}, {Packham}, {Levenson},
  {Elitzur}, {Schaefer}, {Imanishi}, {Sparks}, \& {Radomski}}]{Perlman07}
{Perlman}, E.~S., {Mason}, R.~E., {Packham}, C., {et~al.} 2007, \apj, 663, 808

\bibitem[{{Perola} {et~al.}(2004){Perola}, {Puccetti}, {Fiore}, {Sacchi},
  {Brusa}, {Cocchia}, {Baldi}, {Carangelo}, {Ciliegi}, {Comastri}, {La Franca},
  {Maiolino}, {Matt}, {Mignoli}, {Molendi}, \& {Vignali}}]{Perola04}
{Perola}, G.~C., {Puccetti}, S., {Fiore}, F., {et~al.} 2004, \aap, 421, 491

\bibitem[{{Pier} \& {Krolik}(1992)}]{PK92}
{Pier}, E.~A. \& {Krolik}, J.~H. 1992, \apj, 401, 99

\bibitem[{{Pier} \& {Krolik}(1993)}]{PK93}
{Pier}, E.~A. \& {Krolik}, J.~H. 1993, \apj, 418, 673

\bibitem[{{Risaliti} {et~al.}(2007){Risaliti}, {Elvis}, {Fabbiano}, {Baldi},
  {Zezas}, \& {Salvati}}]{Risaliti07}
{Risaliti}, G., {Elvis}, M., {Fabbiano}, G., {et~al.} 2007, \apjl, 659, L111

\bibitem[{{Risaliti} {et~al.}(2002){Risaliti}, {Elvis}, \&
  {Nicastro}}]{Risaliti02}
{Risaliti}, G., {Elvis}, M., \& {Nicastro}, F. 2002, \apj, 571, 234

\bibitem[{{Risaliti} {et~al.}(1999){Risaliti}, {Maiolino}, \&
  {Salvati}}]{Risaliti99}
{Risaliti}, G., {Maiolino}, R., \& {Salvati}, M. 1999, \apj, 522, 157

\bibitem[{{Schinnerer} {et~al.}(2000){Schinnerer}, {Eckart}, {Tacconi},
  {Genzel}, \& {Downes}}]{Schinnerer00}
{Schinnerer}, E., {Eckart}, A., {Tacconi}, L.~J., {Genzel}, R., \& {Downes}, D.
  2000, \apj, 533, 850

\bibitem[{{Schmitt} {et~al.}(2001){Schmitt}, {Antonucci}, {Ulvestad}, {Kinney},
  {Clarke}, \& {Pringle}}]{Schmitt01}
{Schmitt}, H.~R., {Antonucci}, R.~R.~J., {Ulvestad}, J.~S., {et~al.} 2001,
  \apj, 555, 663

\bibitem[{{Schmitt} {et~al.}(2002){Schmitt}, {Pringle}, {Clarke}, \&
  {Kinney}}]{Schmitt02}
{Schmitt}, H.~R., {Pringle}, J.~E., {Clarke}, C.~J., \& {Kinney}, A.~L. 2002,
  \apj, 575, 150

\bibitem[{{Severgnini} {et~al.}(2003){Severgnini}, {Caccianiga}, {Braito},
  {Della Ceca}, {Maccacaro}, {Wolter}, {Sekiguchi}, {Sasaki}, {Yoshida},
  {Akiyama}, {Watson}, {Barcons}, {Carrera}, {Pietsch}, \&
  {Webb}}]{Severgnini03}
{Severgnini}, P., {Caccianiga}, A., {Braito}, V., {et~al.} 2003, \aap, 406, 483

\bibitem[{{Shlosman} {et~al.}(1990){Shlosman}, {Begelman}, \&
  {Frank}}]{Shlosman90}
{Shlosman}, I., {Begelman}, M.~C., \& {Frank}, J. 1990, \nat, 345, 679

\bibitem[{{Sikora} {et~al.}(2007){Sikora}, {Stawarz}, \& {Lasota}}]{Sikora07}
{Sikora}, M., {Stawarz}, {\L}., \& {Lasota}, J.-P. 2007, \apj, 658, 815

\bibitem[{{Silva} {et~al.}(2004){Silva}, {Maiolino}, \& {Granato}}]{Silva04}
{Silva}, L., {Maiolino}, R., \& {Granato}, G.~L. 2004, \mnras, 355, 973

\bibitem[{{Silverman} {et~al.}(2005){Silverman}, {Green}, {Barkhouse}, {Kim},
  {Aldcroft}, {Cameron}, {Wilkes}, {Mossman}, {Ghosh}, {Tananbaum}, {Smith},
  {Smith}, {Smith}, {Foltz}, {Wik}, \& {Jannuzi}}]{Silverman05}
{Silverman}, J.~D., {Green}, P.~J., {Barkhouse}, W.~A., {et~al.} 2005, \apj,
  618, 123

\bibitem[{{Simpson}(2005)}]{Simpson05}
{Simpson}, C. 2005, \mnras, 360, 565

\bibitem[{{Sirocky} {et~al.}(2008){Sirocky}, {Levenson}, {Elitzur}, {Spoon}, \&
  {Armus}}]{Sirocky08}
{Sirocky}, M.~M., {Levenson}, N.~A., {Elitzur}, M., {Spoon}, H.~W.~W., \&
  {Armus}, L. 2008, \apj, 678, 729

\bibitem[{{Smith} \& {Done}(1996)}]{Smith96}
{Smith}, D.~A. \& {Done}, C. 1996, \mnras, 280, 355

\bibitem[{{Sofue} {et~al.}(1999){Sofue}, {Tutui}, {Honma}, {Tomita},
  {Takamiya}, {Koda}, \& {Takeda}}]{Sofue99}
{Sofue}, Y., {Tutui}, Y., {Honma}, M., {et~al.} 1999, \apj, 523, 136

\bibitem[{{Sparke} \& {Gallagher}(2006)}]{Sparke06}
{Sparke}, L.~S. \& {Gallagher}, III, J.~S. 2006, {Galaxies in the Universe, 2nd
  Edition} (Cambridge University Press)

\bibitem[{{Spoon} {et~al.}(2007){Spoon}, {Marshall}, {Houck}, {Elitzur}, {Hao},
  {Armus}, {Brandl}, \& {Charmandaris}}]{Spoon07}
{Spoon}, H.~W.~W., {Marshall}, J.~A., {Houck}, J.~R., {et~al.} 2007, \apjl,
  654, L49

\bibitem[{{Sturm} {et~al.}(2006){Sturm}, {Hasinger}, {Lehmann}, {Mainieri},
  {Genzel}, {Lehnert}, {Lutz}, \& {Tacconi}}]{Sturm06}
{Sturm}, E., {Hasinger}, G., {Lehmann}, I., {et~al.} 2006, \apj, 642, 81

\bibitem[{{Suganuma} {et~al.}(2006){Suganuma}, {Yoshii}, {Kobayashi},
  {Minezaki}, {Enya}, {Tomita}, {Aoki}, {Koshida}, \& {Peterson}}]{Suganuma06}
{Suganuma}, M., {Yoshii}, Y., {Kobayashi}, Y., {et~al.} 2006, \apj, 639, 46

\bibitem[{{Tristram} {et~al.}(2007){Tristram}, {Meisenheimer}, {Jaffe},
  {Schartmann}, {Rix}, {Leinert}, {Morel}, {Wittkowski}, {R{\"o}ttgering},
  {Perrin}, {Lopez}, {Raban}, {Cotton}, {Graser}, {Paresce}, \&
  {Henning}}]{Tristram07}
{Tristram}, K.~R.~W., {Meisenheimer}, K., {Jaffe}, W., {et~al.} 2007, \aap,
  474, 837

\bibitem[{{Turner} {et~al.}(1997{\natexlab{a}}){Turner}, {George}, {Nandra}, \&
  {Mushotzky}}]{Turner97b}
{Turner}, T.~J., {George}, I.~M., {Nandra}, K., \& {Mushotzky}, R.~F.
  1997{\natexlab{a}}, \apjs, 113, 23

\bibitem[{{Turner} {et~al.}(1997{\natexlab{b}}){Turner}, {George}, {Nandra}, \&
  {Mushotzky}}]{Turner97a}
{Turner}, T.~J., {George}, I.~M., {Nandra}, K., \& {Mushotzky}, R.~F.
  1997{\natexlab{b}}, \apj, 488, 164

\bibitem[{{Urry} \& {Padovani}(1995)}]{Urry95}
{Urry}, C.~M. \& {Padovani}, P. 1995, \pasp, 107, 803

\bibitem[{{Weigelt} {et~al.}(2004){Weigelt}, {Wittkowski}, {Balega}, {Beckert},
  {Duschl}, {Hofmann}, {Men'shchikov}, \& {Schertl}}]{Weigelt04}
{Weigelt}, G., {Wittkowski}, M., {Balega}, Y.~Y., {et~al.} 2004, \aap, 425, 77

\bibitem[{{Whysong} \& {Antonucci}(2004)}]{Whys04}
{Whysong}, D. \& {Antonucci}, R. 2004, \apj, 602, 116

\bibitem[{{Wolf}(2003)}]{Wolf03}
{Wolf}, S. 2003, \apj, 582, 859

\bibitem[{{Yuan}(2007)}]{Yuan07}
{Yuan}, F. 2007, in ASP Conf. Ser. 373: The Central Engine of Active Galactic
  Nuclei, ed. L.~C. {Ho} \& J.-W. {Wang}, 95

\end{thebibliography}

\end{document}